\documentclass[review]{elsarticle}

\usepackage{lineno,hyperref,makecell,amssymb,amsmath}
\usepackage{amstext,amsfonts}
\usepackage{graphicx,subfigure}
\usepackage{caption}
\usepackage{subfigure}
\usepackage{color}
\usepackage{pdfpages}
\usepackage{geometry}
\usepackage{float}
\usepackage{multirow}
\usepackage{longtable}
\usepackage{array}
\newcolumntype{R}[1]{>{\raggedleft\let\newline\\\arraybackslash\hspace{0pt}}m{#1}}
\newcolumntype{L}[1]{>{\raggedright\let\newline\\\arraybackslash\hspace{0pt}}m{#1}}
\newcolumntype{C}[1]{>{\centering\let\newline\\\arraybackslash\hspace{0pt}}m{#1}}

\begin{document}
\captionsetup[figure]{labelfont={bf},name={Fig.},labelsep=period}
\begin{frontmatter}

\title{The performance of SiPM-based gamma-ray detector (GRD) of GECAM-C}


\cortext[cor1]{Corresponding author.}
\author[1]{Dali Zhang\corref{cor1}}
\ead{zhangdl@ihep.ac.cn}

\author[a,b]{Chao Zheng}
\author[a,b]{Jiacong Liu}

\author[1]{Zhenghua An\corref{cor1}}
\ead{anzh@ihep.ac.cn}

\author[a,b]{Chenwei Wang}

\author[1]{Xiangyang Wen\corref{cor1}}
\ead{wenxy@ihep.ac.cn}

\author[a]{Xinqiao Li}
\author[a]{Xilei Sun}

\author[a]{Ke Gong}
\author[a]{Yaqing Liu}
\author[a]{Xiaojing Liu}
\author[a]{Sheng Yang}
\author[a]{Wenxi Peng}
\author[a]{Rui Qiao}
\author[a]{Dongya Guo}
\author[a,b]{Peiyi Feng}
\author[a,b]{Yanqiu Zhang}
\author[a,b]{Wangchen Xue}
\author[a,b]{Wenjun Tan}
\author[a,b,c]{Ce Cai}
\author[a,b,d,e]{Shuo Xiao}
\author[a,f]{Qibin Yi}
\author[a]{Yanbing Xu}
\author[a]{Min Gao}
\author[a]{Jinzhou Wang}
\author[a]{Dongjie Hou}
\author[a]{Yue Huang}
\author[a]{Xiaoyun Zhao}
\author[a]{Xiang Ma}
\author[a]{Ping Wang}
\author[a]{Jin Wang}
\author[a]{Xiaobo Li}
\author[a]{Peng Zhang}
\author[a]{Zhen Zhang}
\author[a]{Yanguo Li}
\author[a]{Hui Wang}
\author[a]{Xiaohua Liang}
\author[a]{Yuxi Wang}
\author[a]{Bing Li}
\author[a]{Jianying Ye}
\author[a]{Shijie Zheng}
\author[a]{Liming Song}
\author[a]{Fan Zhang}
\author[a]{Gang Chen}
\author[a]{Shaolin Xiong}

\address[a]{Key Laboratory of Particle Astrophysics, Institute of High Energy Physics, Chinese Academy of Sciences, Beijing 100049, China}
\address[b]{University of Chinese Academy of Sciences, Beijing 100049, China}
\address[c]{College of Physics, Hebei Normal University, 20 South Erhuan Road, Shijiazhuang 050024, Hebei, China}
\address[d]{Guizhou Provincial Key Laboratory of Radio Astronomy and Data Processing, Guizhou Normal University, Guiyang 550001, GuiZhou, China}
\address[e]{School of Physics and Electronic Science, Guizhou Normal University, Guiyang 550001, GuiZhou, China}
\address[f]{School of Physics and Optoelectronics, Xiangtan University, Xiangtan 411105, Hunan, China}


\begin{abstract}
As a new member of GECAM mission, the GECAM-C (also called High Energy Burst Searcher, HEBS) is a gamma-ray all-sky monitor onboard SATech-01 satellite, which was launched on July 27$^{th}$, 2022 to detect gamma-ray transients from 6 keV to 6 MeV, such as Gamma-Ray Bursts (GRBs), high energy counterpart of Gravitational Waves (GWs) and Fast Radio Bursts (FRBs), and Soft Gamma-ray Repeaters (SGRs). Together with GECAM-A and GECAM-B launched in December 2020, GECAM-C will greatly improve the monitoring coverage, localization, as well as temporal and spectral measurements of gamma-ray transients. GECAM-C employs 12 SiPM-based Gamma-Ray Detectors (GRDs) to detect gamma-ray transients. In this paper, we firstly give a brief description of the design of GECAM-C GRDs, and then focus on the on-ground tests and in-flight performance of GRDs. We also did the comparison study of the SiPM in-flight performance between GECAM-C and GECAM-B. The results show GECAM-C GRD works as expected and is ready to make scientific observations.

\end{abstract}

\begin{keyword}
Gravitational wave \sep GECAM-C \sep SiPM array \sep Gamma-ray detector \sep SiPM array \sep LaBr$_{3}$ detector \sep NaI(TI) detector

\PACS 07.87.+v,29.40.Gx,07.05.Hd
\end{keyword}

\end{frontmatter}

\section{Introduction}
\par The search for gamma-ray bursts (GRBs) coincident with gravitational wave detected by LIGO \cite{LIGO}, Virgo \cite{Virgo} and KAGRA \cite{KAGRA} is critically important in multi-messenger astronomy \cite{Multi-messenger}. Motivated by the first GW-associated GRB170817A event \cite{GW170817}, a series of gamma-ray transient monitors have been proposed, including GECAM \cite{GECAM}, GRID \cite{GRID}, MERGR \cite{MERGer}, BurstCube \cite{BurstCube}, SIRI-1 \cite{SIRI} and so on. Among these projects, SiPM-based gamma-ray detectors are widely adopted for its merits of compact size and low operating voltage.
\par The GECAM mission is a dedicated all-sky gamma-ray monitor to detect and localize gamma-ray transients especially those associated with the gravitational wave events (GWs) and Fast Radio Bursts (FRBs). There are 25 SiPM-based gamma-ray detectors equipped for each GECAM-A and GECAM-B satellite. GECAM-B has been successfully detected many gamma-ray transients, some of which have been reported publicly \cite{GECAMObservation}, while GECAM-A has not been able to regularly observe due to the power supply issues of the satellite platform, thus it is desirable to launch a new GECAM-style monitor to improve the observation capability of GECAM mission.
\par
\begin{figure}[htbp]
  \centering
  \includegraphics[width=6 cm]{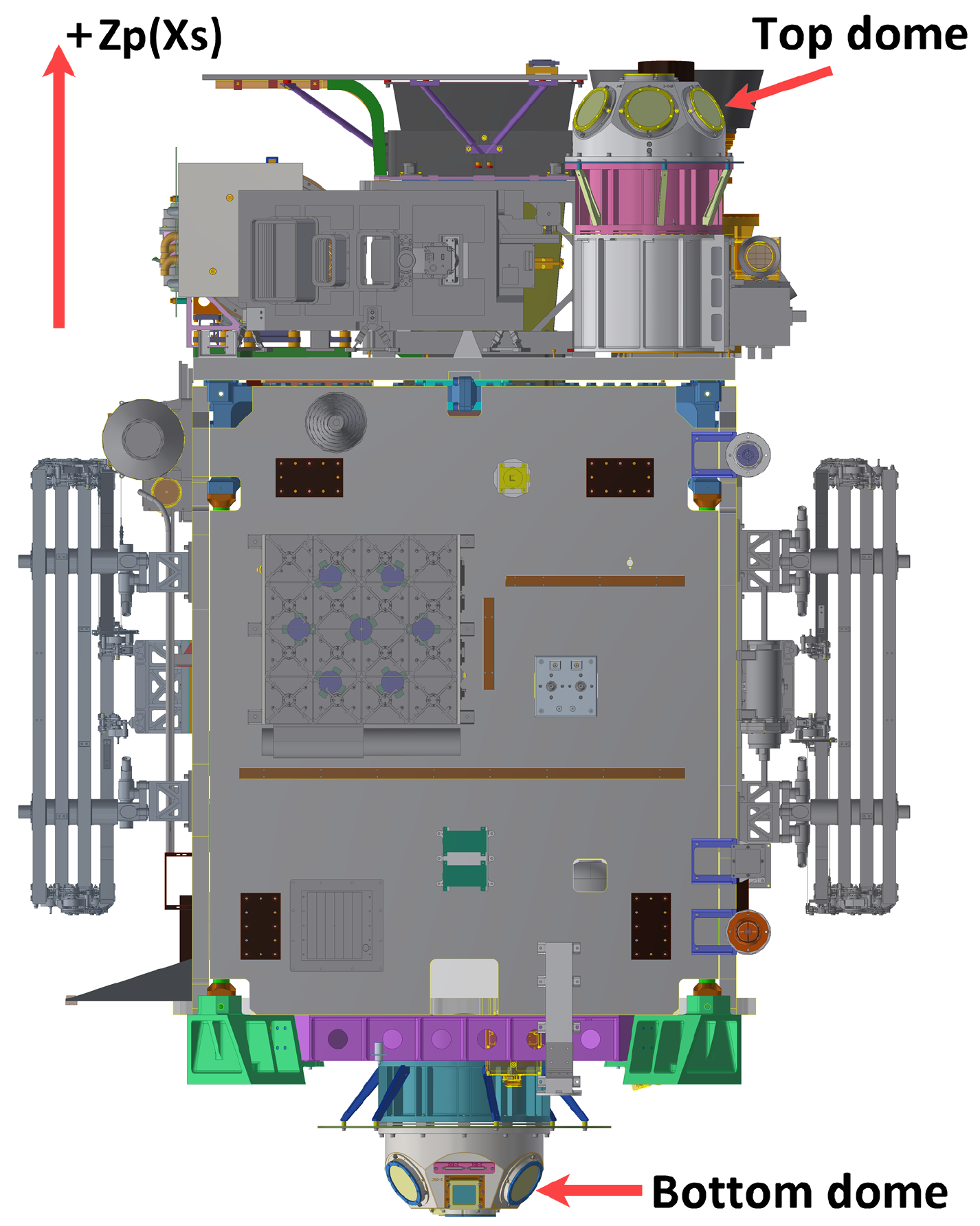}
  \caption{Overview of SATech-01 satellite. GECAM-C consists of two detector domes installed on the top and bottom of the satellite, to achieve a field of view of all-sky occulted by the Earth. The top dome points upwards (+Zp) while bottom dome downwards (-Zp). Subscript p and s indicate the payload and satellite coordinate, respectively. The relation between payload and satellite coordinate is +Xp = +Ys, +Yp = +Zs, +Zp = +Xs.}\label{Fig1}
  \vspace{0.03cm}
\end{figure}

 In 2021, GECAM team proposed the GECAM-C (or High Energy Burst Searcher, HEBS) project which is built mostly based on the GECAM flight-spare detectors and electronics and get launched onboard the Space advanced technology demonstration satellite (SATech-01) \cite{SATech-01} (Fig. \ref{Fig1}) designed by the Innovation Academy for Microsatellites of Chinese Academy of Sciences (IAMCAS). SATech-01 is a sun-synchronous orbit satellite, with an altitude of 500 km and an inclination of 97.4$^{\circ}$. As a new member of GECAM mission, not only the main goal of GECAM-C is the same as GECAM-A and GECAM-B (i.e. to detect and locate X-ray and gamma-ray bursts from 6 keV to 6 MeV, especially those coincident with gravitational waves), but also both the payload technology and scientific application system of GECAM-C are inherited from the GECAM-A and GECAM-B.

\par
\begin{figure}[htbp]
  \centering
  \includegraphics[width=6 cm]{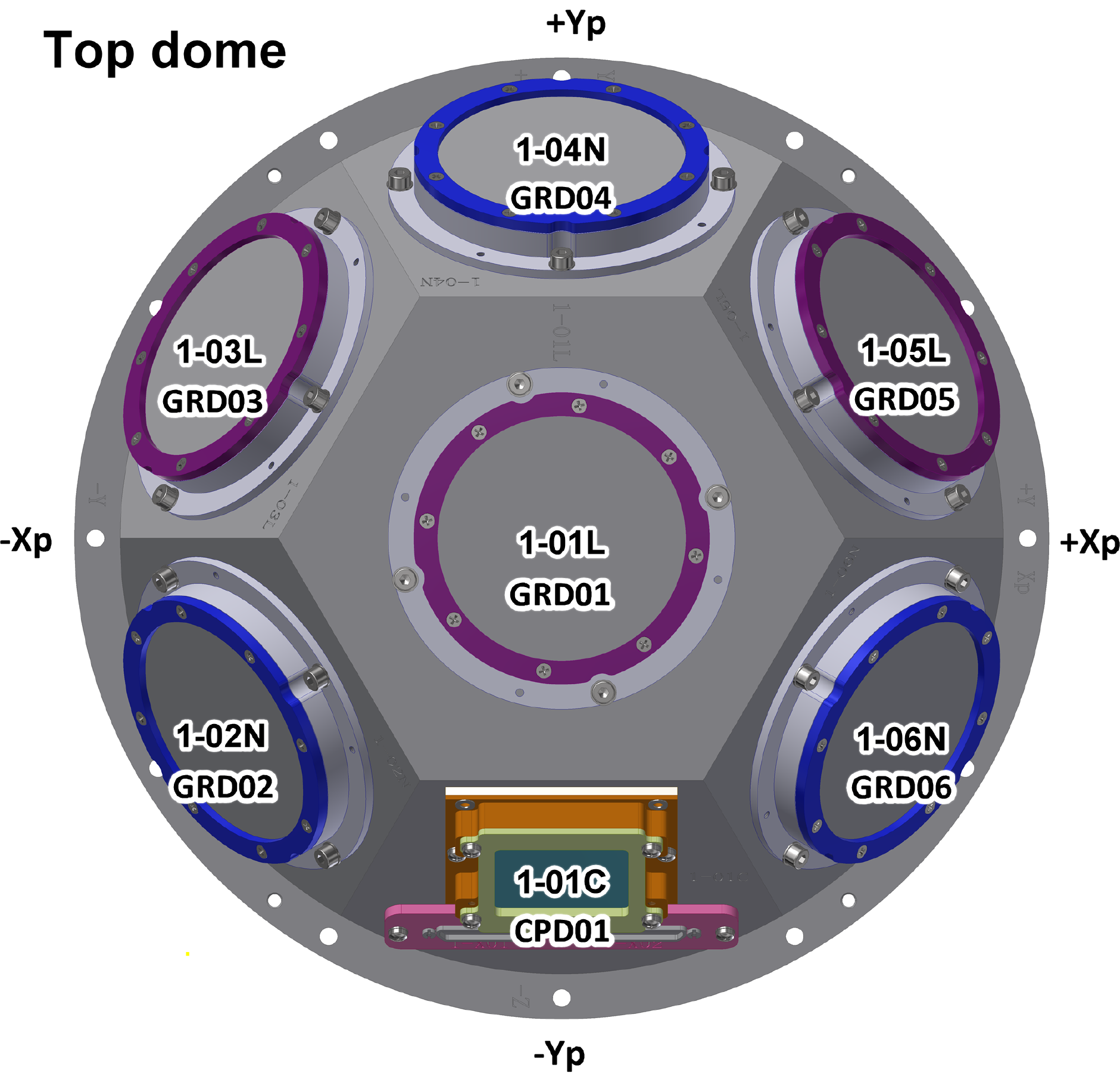}
  \hspace{1cm}
  \includegraphics[width=6.1 cm]{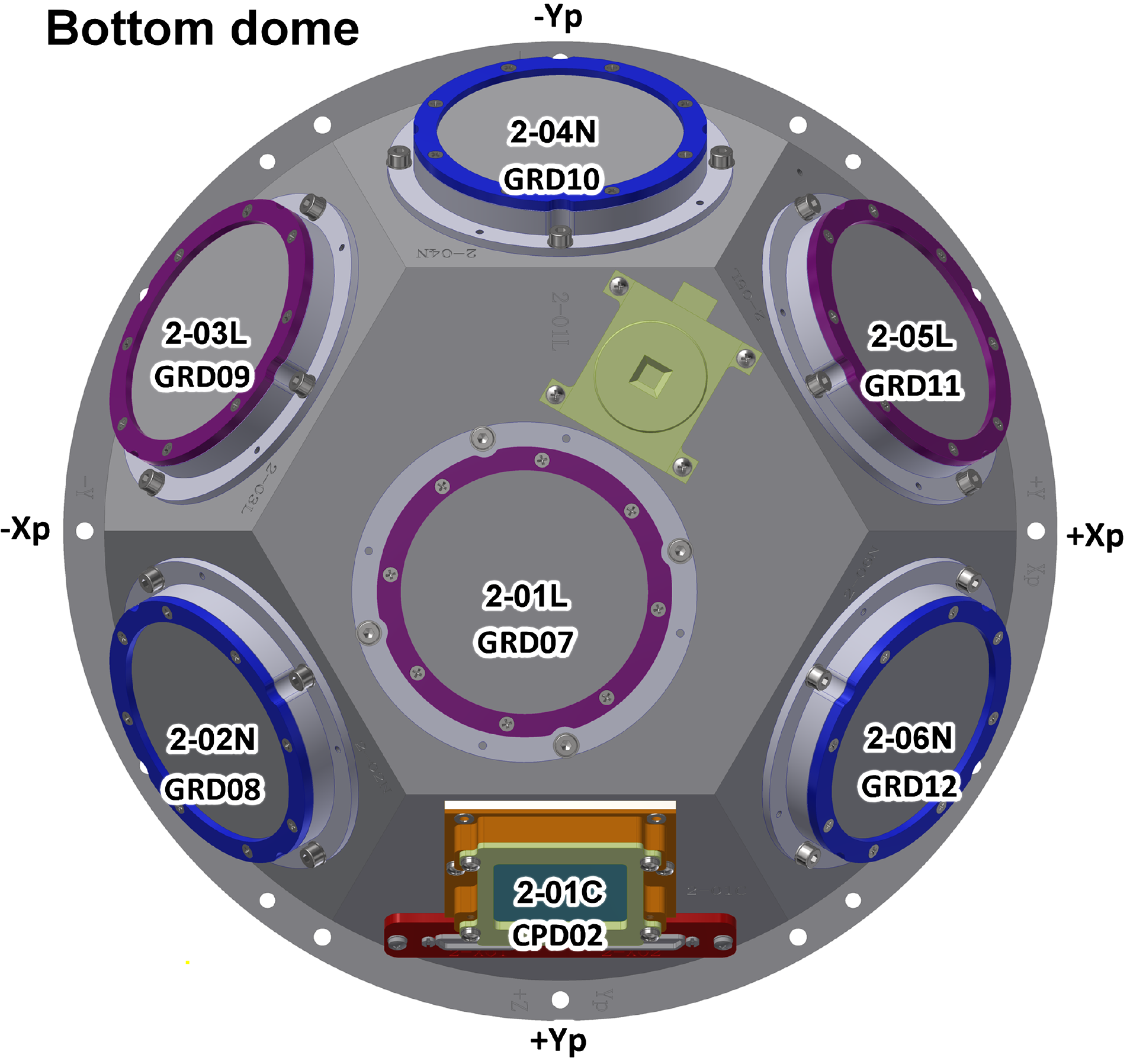}
  \caption{Illustration of GECAM-C domes. There are six (round-shaped) gamma-ray detectors and one (square-shaped) charged particle detector equipped for both the top dome (left panel) and bottom dome (right panel). The assembly ID is labelled. Check Table 2 for the detector ID and type. Subscript p indicates the payload coordinate.}\label{Fig2}
\end{figure}

 As shown in Fig. \ref{Fig2}, GECAM-C payload consists of two detector domes. Each dome contains 3 LaBr$_{3}$ gamma-ray detectors (LGRDs), 3 NaI gamma-ray detectors (NGRDs) and 1 charged particle detector (CPD). Thus, GECAM-C has 12 round-shaped GRDs (i.e. 6 LGRDs and 6 NGRDs) and 2 square-shaped CPDs, pointing to different directions (Fig. \ref{Fig2}). GRD is made with LaBr$_{3}$ and NaI scintillator, and is the main detector to measure the positional, temporal and spectral properties of gamma-ray transients. CPD \cite{CPD} is made by plastic scintillator and it is used to help to discriminate charged particle events and gamma-ray bursts. The scintillation signals of all GRDs and CPDs are readout by SiPM arrays. LGRDs and CPD are flight-spare detectors of GECAM-A and GECAM-B. To save cost and test new technologies, NGRDs are also developed and used in GECAM-C.
\par The localization accuracy and field of view can be greatly improved through joint observation of GECAM-B, GECAM-C, Insight-HMXT and other in-flight transient source monitors \cite{GRB210121A} \cite{GECAMEnhancedLocalization} \cite{GECAMJointAnalysis}. GECAM-C also has the real-time GRB trigger system based on BeiDou short message that has been successfully applied in GECAM-B \cite{GECAMTrigger}. Real-time GRB trigger alerts will help follow-up observations by other high energy telescopes or in other electromagnetic wavelength.
\par GECAM-C was successfully launched on July 27$^{th}$, 2022 (Beijing Time). The commissioning observation starts from Aug 4$^{th}$, 2022. Since then, GECAM-C has been routinely detecting gamma-ray transients (gamma-ray bursts, soft gamma-ray repeater, solar flares, etc.) and sending the near real-time trigger alerts. During the commissioning phase, GECAM-C, together with GECAM-B, also discovered the second x-ray burst associated with radio burst from SGR J1935+2154 \cite{ATel15682}. GECAM-C detected the prompt emission of the extraordinary bright burst GRB 221009A and made an unprecedented measurement of the most bright part without data saturation \cite{HEBSGRB221009A}.
\par  In this paper, we present the design, energy gain calibration and temperature dependence tests of GECAM-C GRD. The in-flight GRD performance during the commissioning phase is also displayed.
\section{GRD design}
\par
\begin{figure}[htbp]
  \centering
  \includegraphics[width=6 cm]{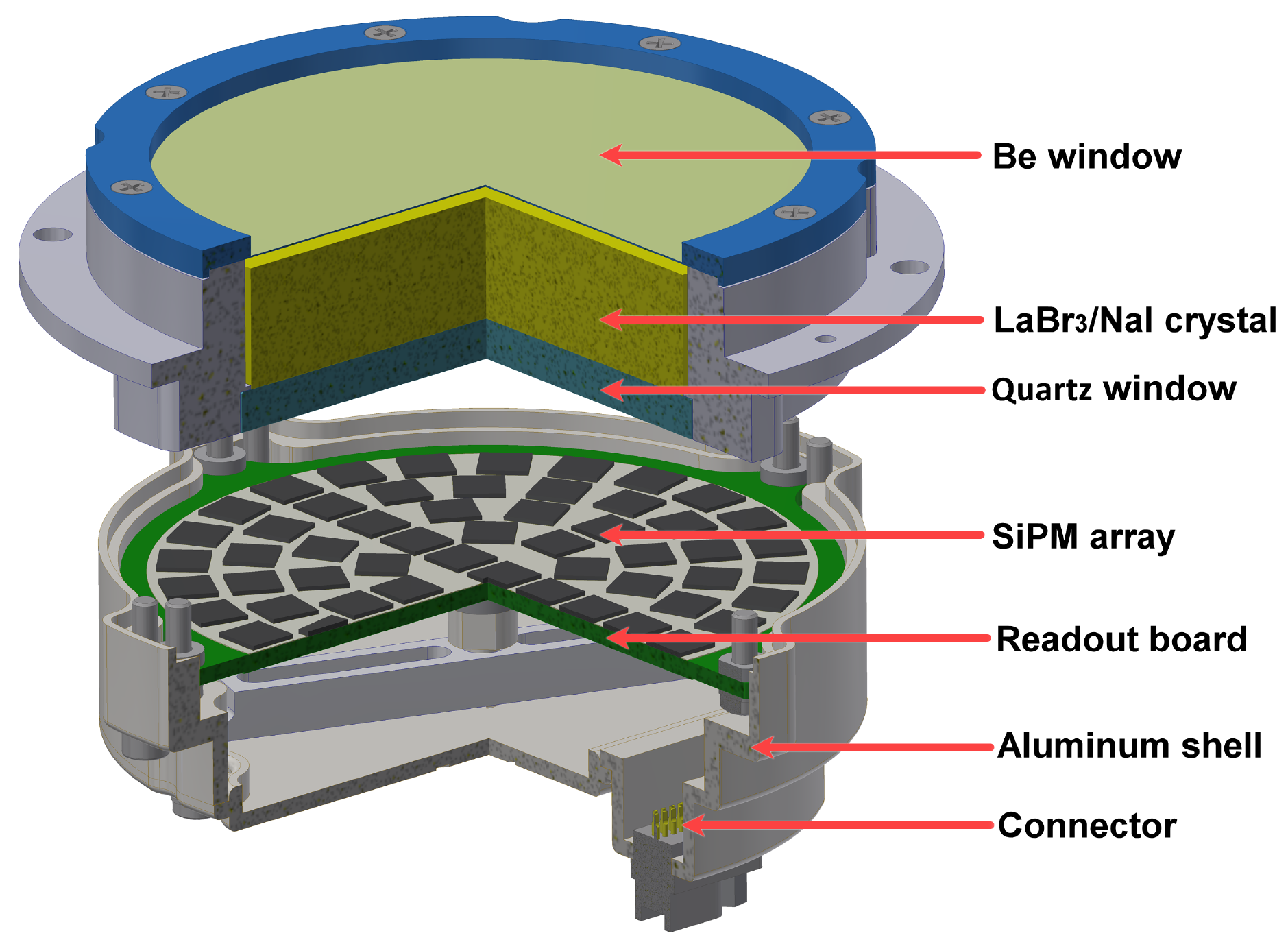}
  \caption{Illustration of the GECAM-C GRD detector.}\label{Fig3}
  \vspace{0.05cm}
\end{figure}

 As shown in Fig. \ref{Fig3}, the GECAM-C GRD has similar design with GECAM \cite{GECAMGRD} \cite{GECAMSiPMArray}. The 3-inch scintillator of GRDs are made of LaBr$_{3}$ or NaI(Tl) scintillator with a thickness of 15 mm. All the scintillators are provided by Beijing Glass Research Institute. The incident window is made of a 0.2 mm-thick Be sheet. The light output window is equipped with quartz glass and then coupled with a 64-chips SiPM array (MICROFJ-60035-TSV-TR) through silicone rubber. The SiPM array is placed on the front of the readout circuit board.
\par According to the SiPM application experience of GECAM-B, we reduced the SiPM bias voltage of GECAM-C GRD to reduce the current growth rates caused by irradiation damage. At the same time, the gain of SiPM pre-amplifiers are increased to compensate the decreased SiPM gain under lower bias voltage. Lower operating current of SiPM is expected to enhance the lifetime. An adaptive voltage supply source was also designed to automatically adjust the SiPM bias voltage to compensate SiPM temperature effects \cite{GECAMGain}.
\par
\begin{figure}[htbp]
  \centering
  \setlength{\belowcaptionskip}{0.1cm}
  \includegraphics[width=10 cm]{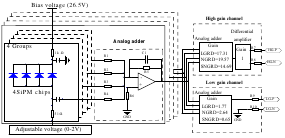}
  \caption{Electronics design of SiPM array and preamplifier.}\label{Fig4}
\end{figure}

 As shown in Fig. \ref{Fig4}, the signal of SiPM array is amplified and shaped by the pre-amplifier on the back of the readout circuit board, and the differential output signals are grouped into high gain channel and low gain channel to achieve a large dynamic range. The high and low gain channels of 6 LGRDs and 4 NGRDs are fed to the data acquisition system \cite{GECAMDataAcquisition}. Due to the limited data acquisition electronic resource, only the high gain channel (rather than both channels) of 2 NGRDs is readout by data acquisition system, so they are called as single-channeled NGRD (SNGRD).
\par To meet the energy range requirements of GECAM-C, the gain of the GRD pre-amplifier are optimized based on the type of GRD (i.e. LGRD, NGRD, SNGRD). The specifications of various GRDs are listed in Table \ref{GRD designed}. Because of the complex in-flight operation conditions, the noise level and energy range of different GRDs are different and may evolve with time. The assembly specifications of GECAM-C detector were listed in Table \ref{GRD assembly}.

\begin{table}[H]
\centering
\caption{Typical characteristics of GRD in-flight performance. }
\begin{tabular}{|c|c|}
\hline
  SiPM bias voltage & 24.5$\backsim$26.5 V\\ \hline
  Working temperature & -20$\pm$5 $^{\circ}$C\\ \hline
  Deadtime & $\leq$4 $\mu s$ (for normal event)  \\ \hline

  \multirow{2}{*}{Energy range of dual channel GRD} & High gain channel: 6$\backsim$300 keV \\ \cline{2-2}
  \multirow{2}{*}{}&Low gain channel: 160 keV$\backsim$6 MeV\\ \hline

  Energy range of single channel NGRD & 6$\backsim$450 keV \\ \hline

  \multirow{2}{*}{Energy resolution} & LGRD: $\approx$ 21\% @ 37.4 keV \\ \cline{2-2}
  \multirow{2}{*}{}& NGRD: $\approx$ 11\% @ 191 keV\\

\hline
\end{tabular}
\label{GRD designed}
\end{table}
\begin{table}[H]
\centering
\caption{ Assembly specifications of GECAM-C detectors. Pointing direction is in
the payload coordinate.}
\begin{tabular}{|c|c|c|c|C{1.5cm}|C{1.5cm}|}
\hline
\multirow{2}{*}{Dome type}           &Assembly      & Detector  & Detector  & \multicolumn{2}{c|}{Incident direction ($^{\circ}$)}  \\ \cline{5-6}
\multirow{2}{*}{}                    &ID	        &ID	        &  type	    & theta  &phi  \\ \cline{2-6}
\multirow{7}{*}{Top dome}	         &1-01L	        &GRD01	    &  LGRD    	&  0     &0    \\ \hline
\multirow{7}{*}{}                    &1-02N	        &GRD02	    &  NGRD    	&  60    &210  \\ \cline{2-6}
\multirow{7}{*}{}                    &1-03L	        &GRD03	    &  LGRD    	&  60    &150  \\ \cline{2-6}
\multirow{7}{*}{}                    &1-04N	        &GRD04	    &  NGRD    	&  60    &90   \\ \cline{2-6}
\multirow{7}{*}{}                    &1-05L	        &GRD05	    &  LGRD    	&  60    &30   \\ \cline{2-6}
\multirow{7}{*}{}                    &1-06N	        &GRD06	    &  SNGRD	&  60    &330  \\ \cline{2-6}
\multirow{7}{*}{}                    &1-01C	        &CPD01	    &  CPD	    &  60    &270  \\ \hline
\multirow{7}{*}{Bottom dome}         &2-01L	        &GRD07	    &  LGRD	    &  180   &0    \\ \cline{2-6}
\multirow{7}{*}{}                    &2-02N         &GRD08	    &  NGRD	    &  120   &150  \\ \cline{2-6}
\multirow{7}{*}{}                    &2-03L	        &GRD09	    &  LGRD	    &  120   &210  \\ \cline{2-6}
\multirow{7}{*}{}                    &2-04N         &GRD10	    &  NGRD	    &  120   &270  \\ \cline{2-6}
\multirow{7}{*}{}                    &2-05L	        &GRD11	    &  LGRD	    &  120   &330  \\ \cline{2-6}
\multirow{7}{*}{}                    &2-06N         &GRD12	    &  SNGRD	&  120   &30   \\ \cline{2-6}
\multirow{7}{*}{}                    &2-01C         &CPD02	    &  CPD	    &  120   &90   \\ \hline
\end{tabular}
\label{GRD assembly}
\end{table}

\section{On-ground test results}
\subsection{Calibration test in Hard X-ray Calibration Facility}
\par
\begin{figure}[htbp]
  \centering
  \setlength{\belowcaptionskip}{0.1cm}
  \includegraphics[width=7 cm]{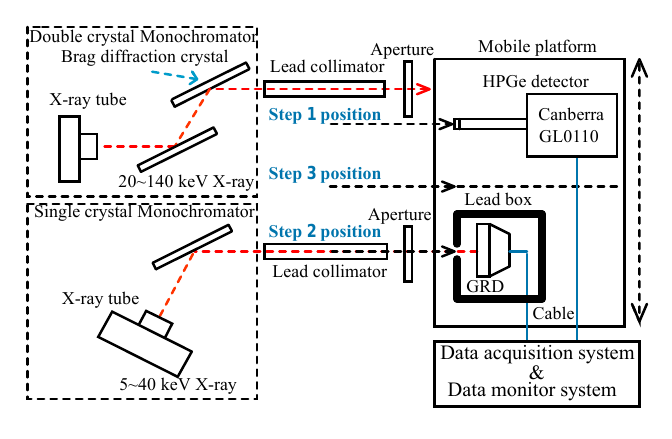}
  \caption{Calibration test setup.}\label{Fig5}
\end{figure}

 The energy response of GRD is characterized with the Hard X-ray Calibration Facility (HXCF) at the National Institute of Metrology \cite{XrayFacility}. As shown in Fig. \ref{Fig5}, the Hard X-ray Calibration Facility consists of double crystal monochromator X-ray beam (20-140 keV) and single crystal monochromator beam (5-40 keV). A mobile platform is used to move the tested detectors to the desired beam position. GRD is housed in a lead box to shield scattered X-rays in the environment. A well-calibrated high-purity germanium (HPGe) detector was used to monitor the beam flux and energy for each test run. The calibration process for each energy includes 3 steps: The first step is to move the mobile platform until the HPGe detector is in the emitting direction of the X-ray beam. HPGe detector measures the energy and total counts of X-ray beam for 100 s as standard calibration data. The second step is to move the mobile platform untill the GRD is in the emitting direction of X-ray beam. GRD will measure the X-ray beam for 100 s. The third step is to move the mobile platform untill the emitting direction of X-ray beam is in the middle position of GRD and HpGe detector. Both of the GRD and HpGe detector will measure the background for 100 s. All GRDs are tested under a fixed SiPM bias voltage of 26.5 V.

\par
\begin{figure}[htbp]
  \centering
  \setlength{\belowcaptionskip}{0.1cm}
  \includegraphics[width=7 cm]{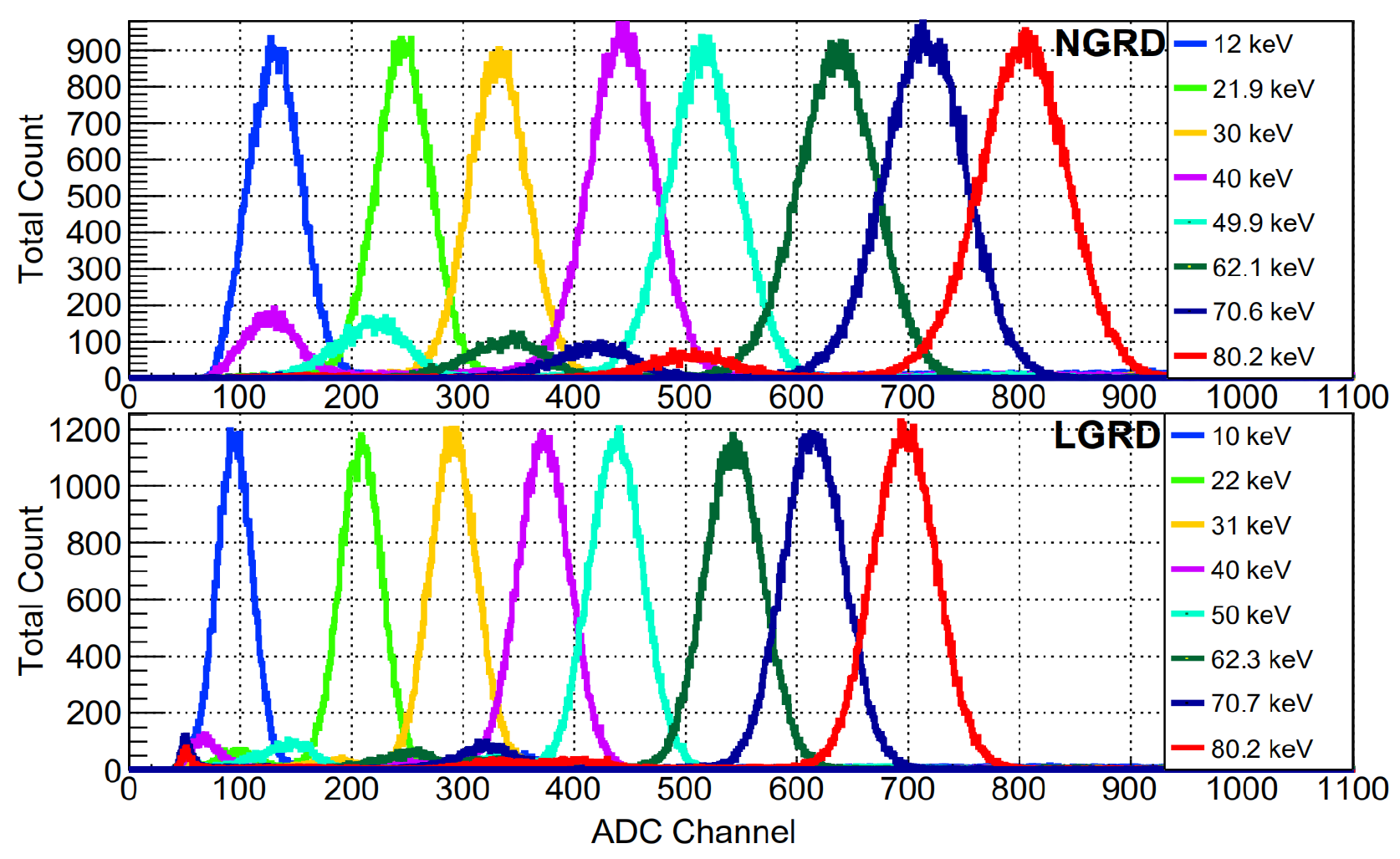}
  \caption{Spectra measured by GRDs for different energy points.}\label{Fig6}
\end{figure}

\begin{figure}[htbp]
  \centering
  \includegraphics[width=7 cm]{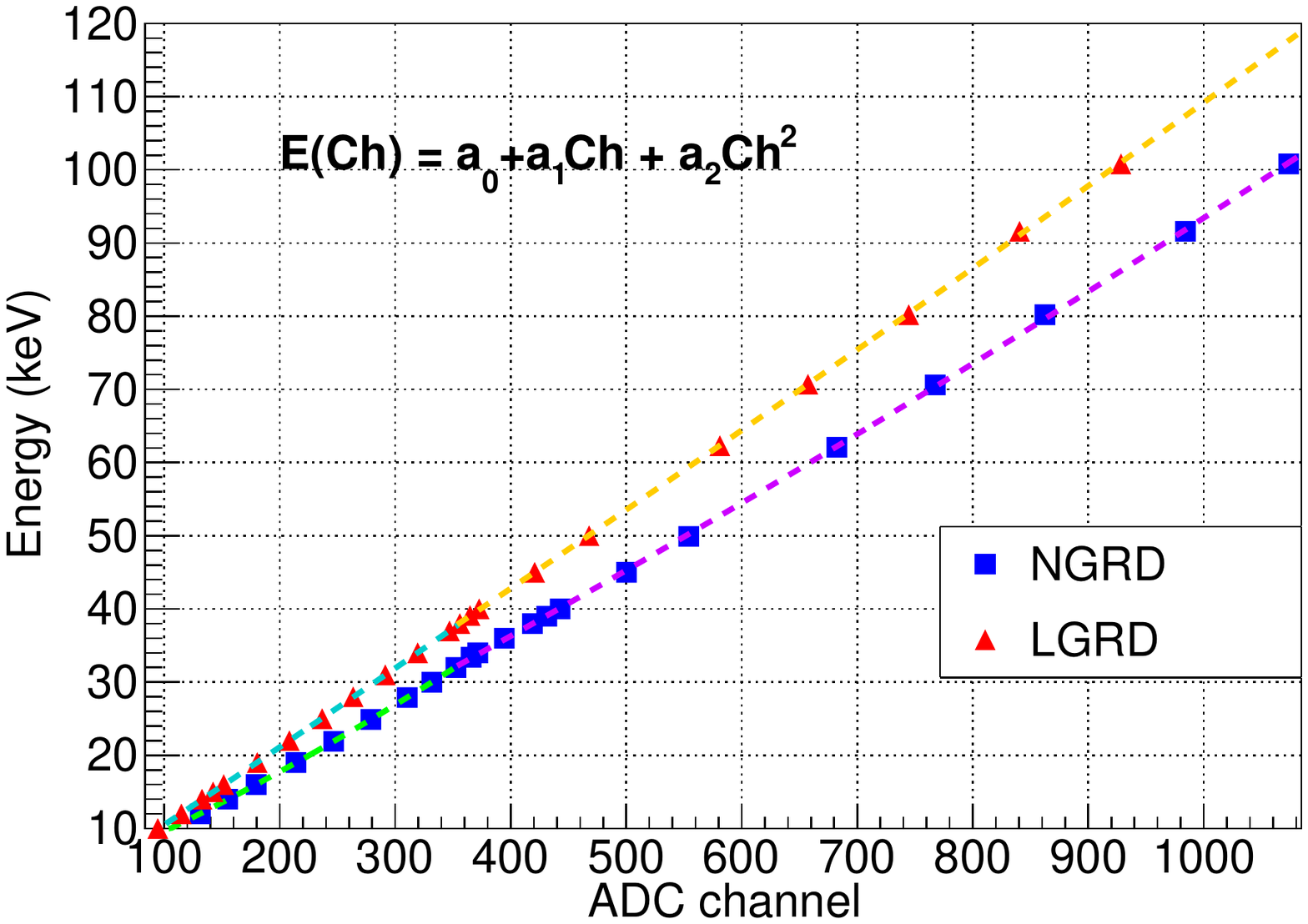}
  \hspace{1cm}
  \includegraphics[width=7 cm]{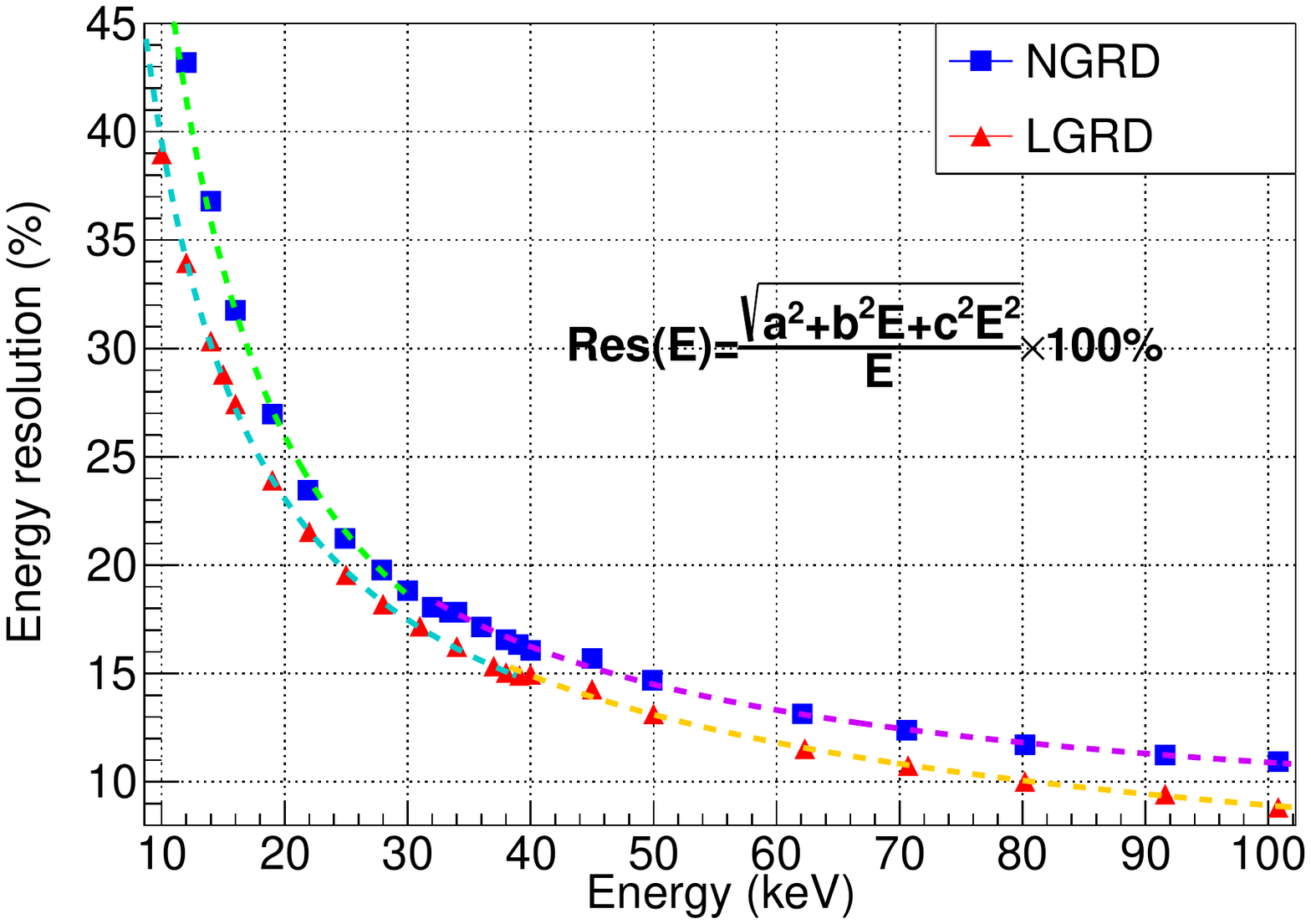}
  \caption{Left panel: E-C relation for LGRD and NGRD of GECAM-C. Right panel: Energy resolution as a function of energy for LGRD and NGRD of GECAM-C.}\label{Fig7}
\end{figure}

 Fig. \ref{Fig6} shows energy spectra measured by GRDs during calibration tests. In the data analysis process, the background is deducted first and a single Gaussian function is used to fit the full-energy peak. The energy peak position and energy resolution are extracted from the fitting results of mean value and $FWHM$ respectively. The total counts of GRD is obtained by accumulating the 2.58 sigma range of the Gaussian distribution. The dead time is recorded in the event by event data and deducted from the total time to get the live time. The measured E-C relation and energy resolution (Energy range 10-100 keV) of LGRD and NGRD are shown in Fig. \ref{Fig7}. Detailed calibration analysis of the X-ray beam and radioactive source tests are presented in a dedicated paper \cite{HEBSCal}. The results of the on-ground calibration test are used as a benchmark of the calibration database. During the integrated assembly test of the GECAM-C payload and the satellite, only a limited number of energy points of radioactive sources are available for energy calibration to cross check the calibration database. After launch, the energy calibration will be based on the in-flight characteristic lines measured by the detectors.

\subsection{Temperature dependence test}
\par In the long term operation of GECAM-C payload, there will be a temperature fluctuation with the satellite orbit. Because SiPM-based GRD has an obvious temperature dependence, it may result in a gain drift of GRD and will cause severe problem for energy response calibration in the data analysis process. We have developed a real-time gain stabilization and consistency correction approach for multiple SiPM-based gamma-ray detectors \cite{GECAMGain}. This approach is based on the adaptive voltage source that updates the SiPM bias voltage according to the flexible temperature–bias-voltage look-up table (LUT). The design of LUT is based on the tested temperature dependence of GRD. Because temperature dependence of LGRD was already well studied in the GECAM project, thus in this section we only show the test results of NaI-based GRD (NGRD) for GECAM-C.
\begin{figure}[htbp]
  \centering
  \includegraphics[width=7.3 cm]{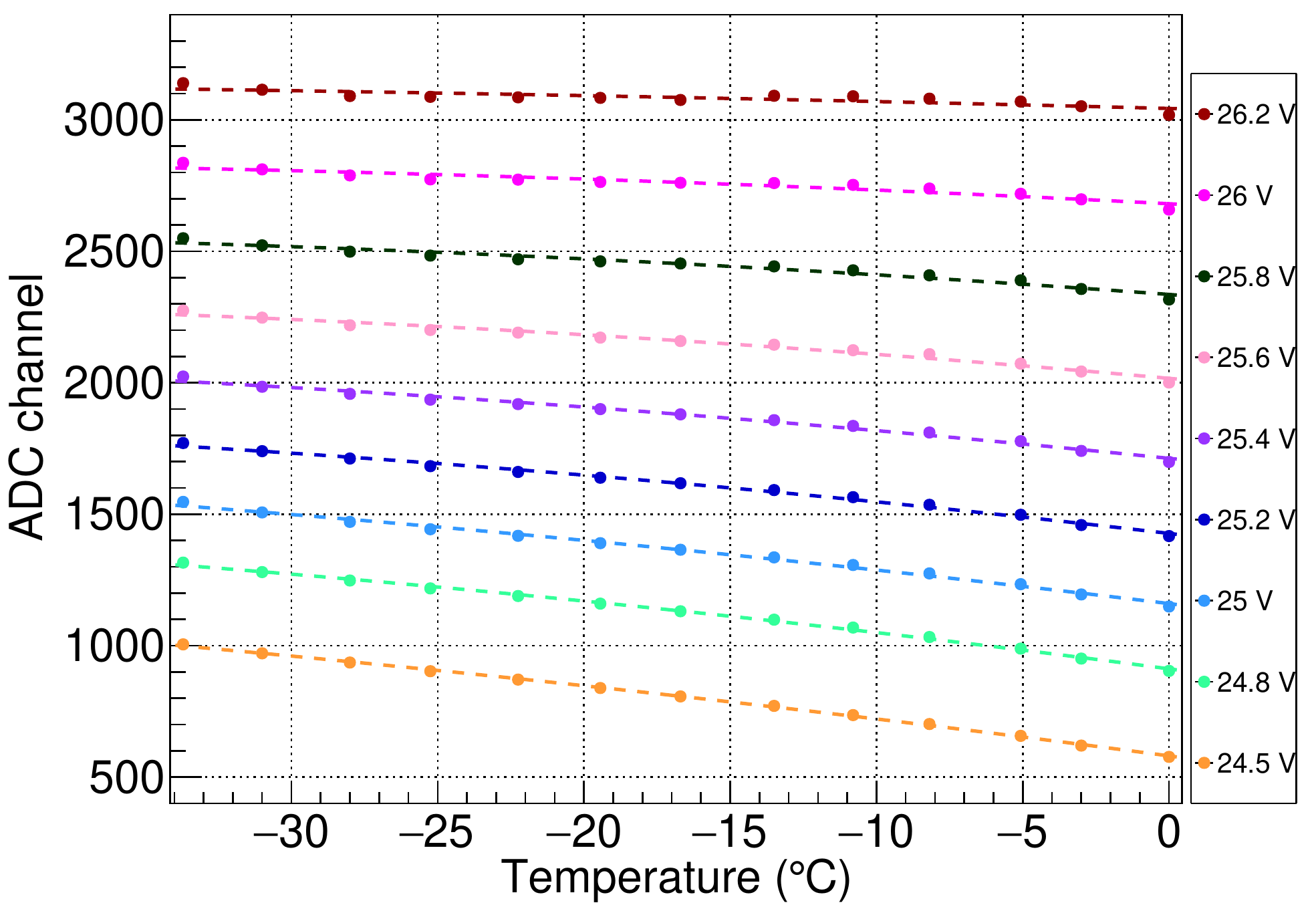}
  \hspace{1cm}
  \includegraphics[width=6.7 cm]{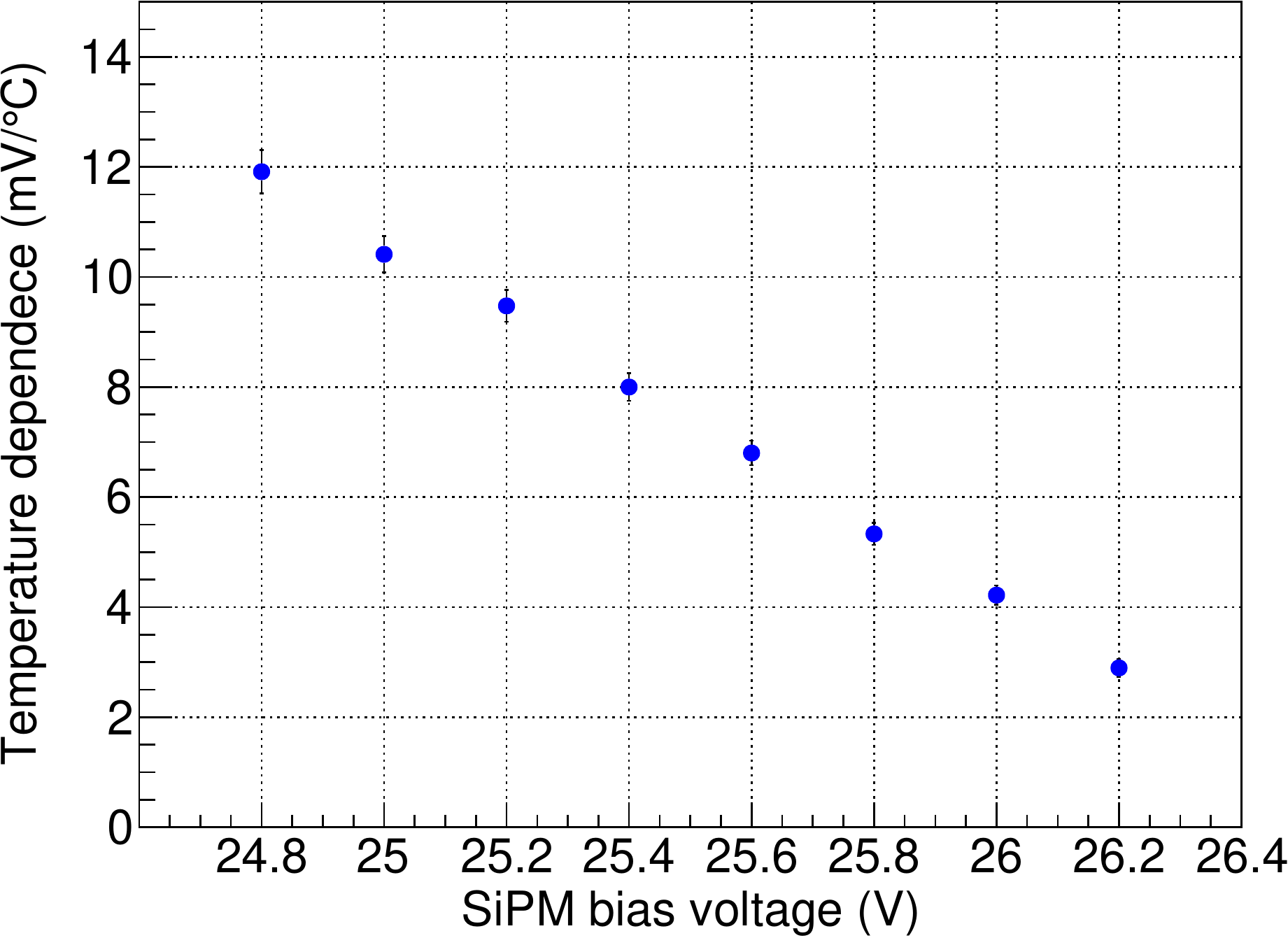}
  \caption{Left panel: On-ground test results of GRD temperature dependence. The peak positions of 511keV under various bias voltage and temperature are measured. Right panel: temperature dependence Vs SiPM bias voltage. }\label{Fig8}
\end{figure}
\par The 511 keV gamma-ray energy line of $^{22}$Na radioactive source was used to measure the gain drift of NGRD within the temperature range of -35 $^{\circ}$C to 0 $^{\circ}$C and bias voltage range of 24.5 V to 26.2 V. Fig. \ref{Fig8} left panel shows the ADC channel (Chn) of the energy line peak will increase with the decrease of temperature (T) which is fit by Chn=a+b·T+c·T$^{2}$. NaI crystal has a negative temperature dependence \cite{NaITemp}, while SiPM has a positive temperature dependence \cite{SiPMTemp}. The test results show NGRD has an overall positive temperature dependence (P$_{td}$). To keep the GRD gain stable, the SiPM bias voltage (V$_{b}$) need to be adjusted with present temperature (T) as V$_{b}$=V$_{T0}$+P$_{td}$·(T-T$_{0}$) , where V$_{T0}$ is the SiPM bias voltage under a reference temperature (T$_{0}$). The overall NGRD temperature dependence (P$_{td}$) has a similar definition with SiPM temperature dependence and it can be calculated as Eq.\ref{Func}: V$_{0}$ is the reference SiPM bias voltage which is selected as the lower and closest bias voltage to V$_{b}$ in the measured data. Chn$_{b}$ is the ADC channel of 511 keV peak position at T$_{b}$ and V$_{b}$. T$_{b}$ is selected as 0$^{\circ}$C and T$_{0}$(V$_{0}$,Chn$_{b}$) is determined from data of Fig. \ref{Fig8} left panel. Fig. \ref{Fig8} right panel shows the tested results of NGRD temperature dependence.
\begin{equation}
P_{td}(V_b) = \frac{V_{b}-V_{0}}{T_{b}(V_{b},Chn_{b})-T_{0}(V_{0},Chn_{b})} =-\frac{V_{b}-V_{0}}{T_{0}(V_{0},Chn_{b})}
\label{Func}
\end{equation}

\par
\begin{figure}[htbp]
  \centering
  \setlength{\belowcaptionskip}{0.1cm}
  \includegraphics[width=8 cm]{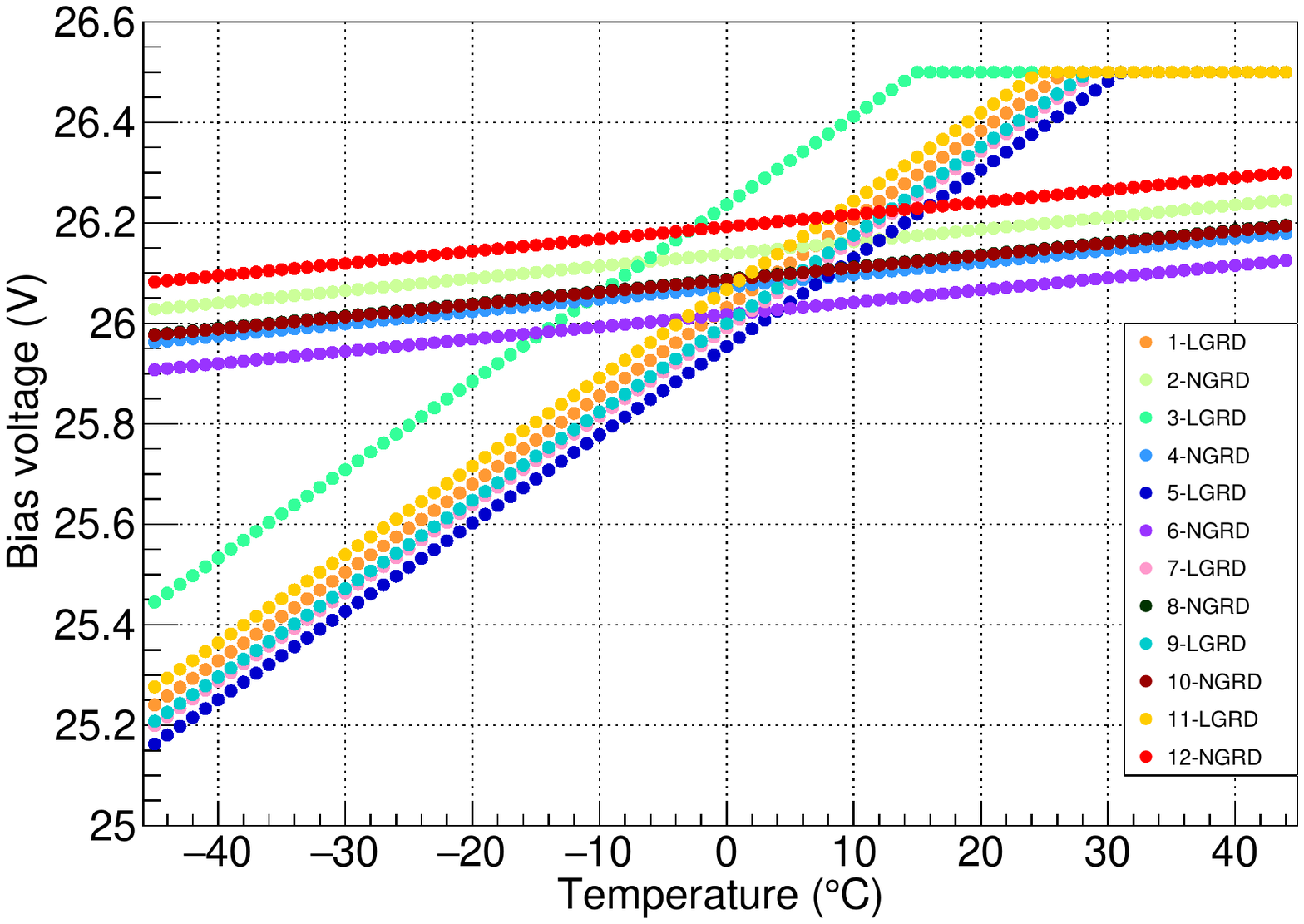}
  \caption{Designed temperature bias look-up table in assembly tests of the flight model in phase D. The adjustable SiPM bias voltage range is 24.5 V to 26.5 V.}\label{Fig9}
\end{figure}

 According to the test results, the bias voltage LUT for GRD is determined. Fig. \ref{Fig9} shows bias voltage LUT of 12 GRDs that covers a temperature range of -45 $^{\circ}$C to 45 $^{\circ}$C. The data acquisition system will scan the temperature monitor on GRD once per second. If temperature changes larger than 0.5 $^{\circ}$C, SiPM bias voltage will be refreshed according to bias voltage LUT. In the on-ground test, the SiPM bias voltage of NGRD is around 26.1 V at 20 $^{\circ}$C, and the corresponding temperature dependence of NGRD is 2.8 mV/$^{\circ}$C. For LGRD, the temperature dependence is set to be 18 mV/ $^{\circ}$C which is already successfully applied in GECAM-A and GECAM-B. The light yield difference of GRDs results in the gain non-uniformity. To obtain better gain consistency, each GRD has its proper SiPM bias voltage at the same temperature as Fig. \ref{Fig9} shows. The overall performance of GRD will deteriorate due to the long term in-orbit irradiation and the evolving temperature dependence. To deal with this situation, the bias voltage LUT is designed to be updatable by command from the ground.
\subsection{Final on-ground test before launch}
\par
\begin{figure}[htbp]
  \centering
  \includegraphics[width=6.7 cm]{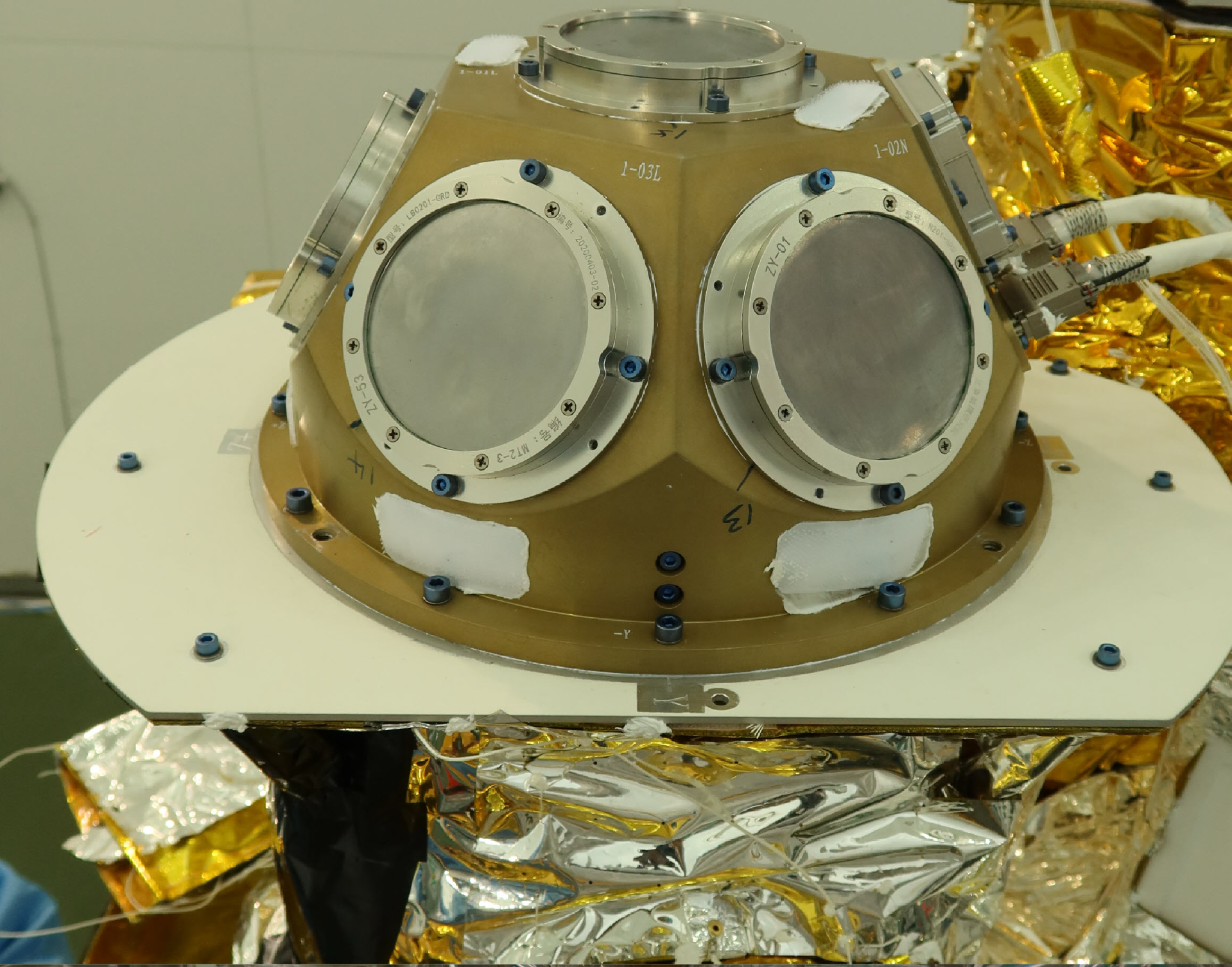}
  \hspace{1cm}
  \includegraphics[width=7.2 cm]{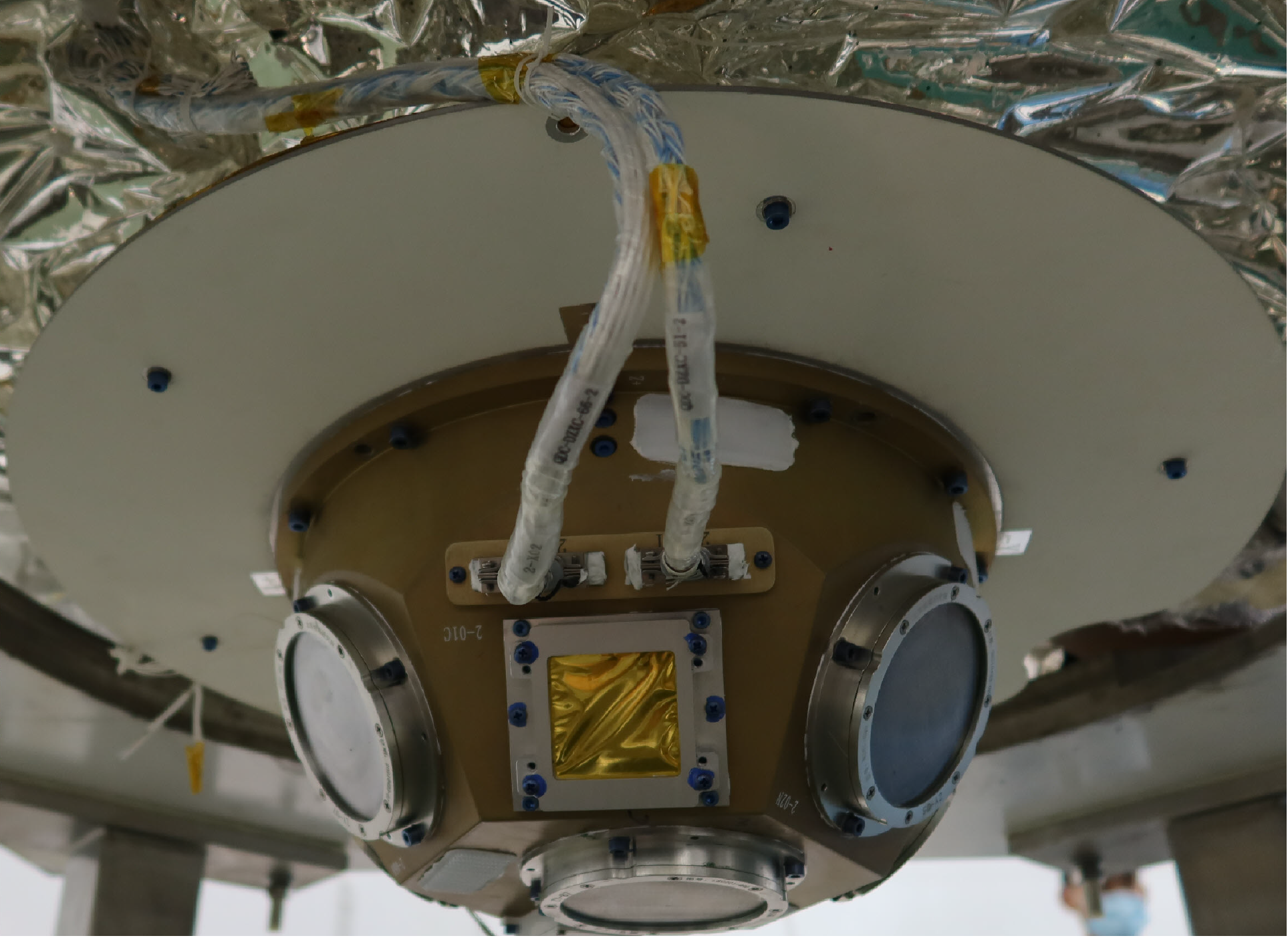}
  \caption{GECAM-C final on-ground calibration test in Jiuquan Satellite Launch Center. Left panel: top dome. Right panel: bottom dome. These photos were taken before the GECAM-C payload was wrapped with thermal insulation. }\label{Fig10}
\end{figure}

\begin{figure}[htbp]
  \centering
  \setlength{\belowcaptionskip}{0.1cm}
  \includegraphics[width=12 cm]{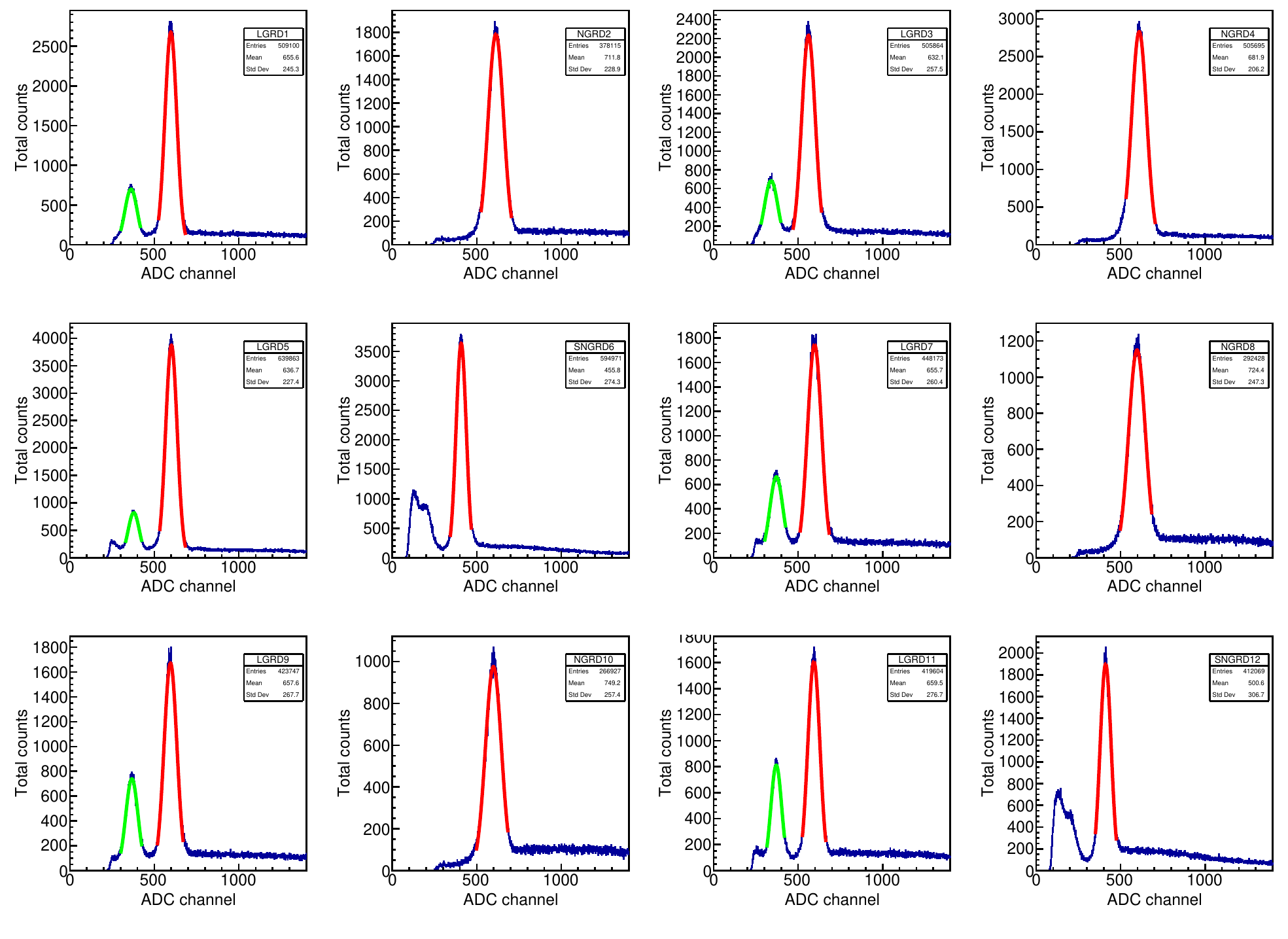}
  \caption{Energy spectra measured in the radioactive source test.}\label{Fig11}
\end{figure}

 After all GRDs were installed on GECAM-C payload and satellite, the final on-ground test was conducted at China's Jiuquan Satellite Launch Center in May 2022. As shown in Fig. \ref{Fig10}, two domes of GECAM-C were installed on the top and bottom sides of the satellite respectively. In calibration test, $^{241}$Am radioactive sources were used to measure the energy response of GRDs. The 59.5 keV peak positions of all GRDs were adjusted by updating the SiPM temperature–bias-voltage LUT by command. As shown in Fig. \ref{Fig11}, all the 59.5 keV energy peaks (red lines) were adjusted to be aligned. In addition to the energy line of $^{241}$Am, LGRD has an intrinsic 37.4 keV peak from $^{138}$La and it is marked with green lines.

\par
\begin{figure}[htbp]
  \centering
  \setlength{\belowcaptionskip}{0.1cm}
  \includegraphics[width=8.5 cm]{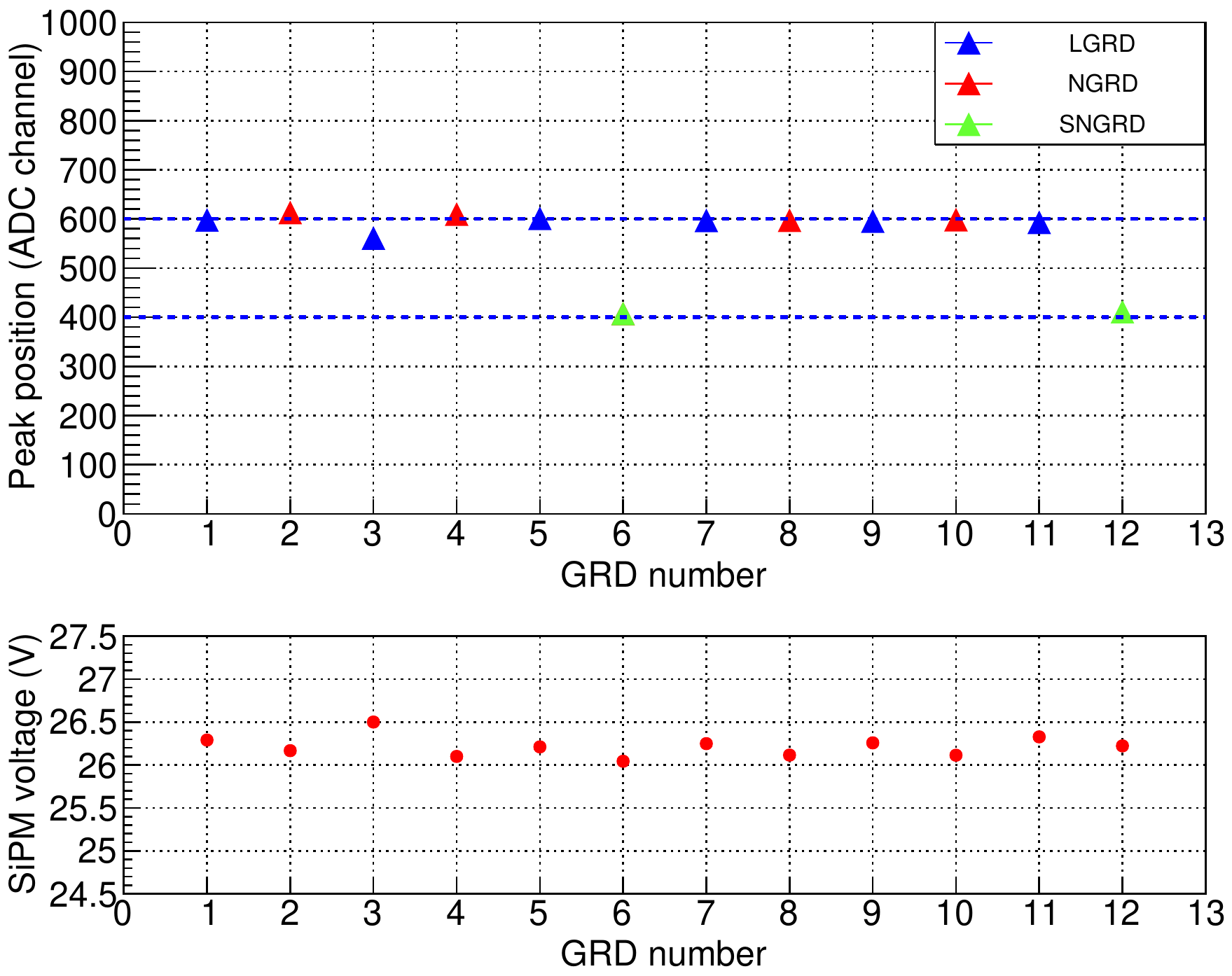}
  \caption{The 59.5 keV peak position of all GRDs after gain adjustment during the final ground test.}\label{Fig12}
\end{figure}

 Fig. \ref{Fig12} shows the results of peak position alignment among detectors. The target position of 59.5 keV peak of NGRD and LGRD is 600 ADC and SNGRDs is 400 ADC, respectively. Each GRD has its own SiPM bias voltage which is determined by the gain consistency correction approach \cite{GECAMGain}. Because GRD03 has a lower light yield, even if the voltage was set to the maximum, the peak position was still lower than other GRDs. However, the SiPM gain will increase at lower temperature in-orbit, so the gain of GRD03 will be adjusted to be consistent with other GRDs during in-flight commissioning phase.
\section{In-flight performance}
\subsection{In-flight background}
\par After launch, the GRD operation strategy has been optimized according to the in-flight situation. The in-flight background of GRD mainly consists of cosmic X-ray background, albedo gamma, cosmic proton, scintillator inner radiation, proton activated radiation \cite{Inflight_BK} and so on. Characteristic energy lines that could be resolved from the continuous background spectrum are used for in-flight energy gain calibration. However, GECAM-C will pass through the high latitude area and South atlantic anomaly (SAA) area frequently and the high flux of proton and electron will severely affect the measurements of the in-flight characteristic energy lines.
\begin{figure}[htbp]
  \centering
  \includegraphics[width=6.7 cm]{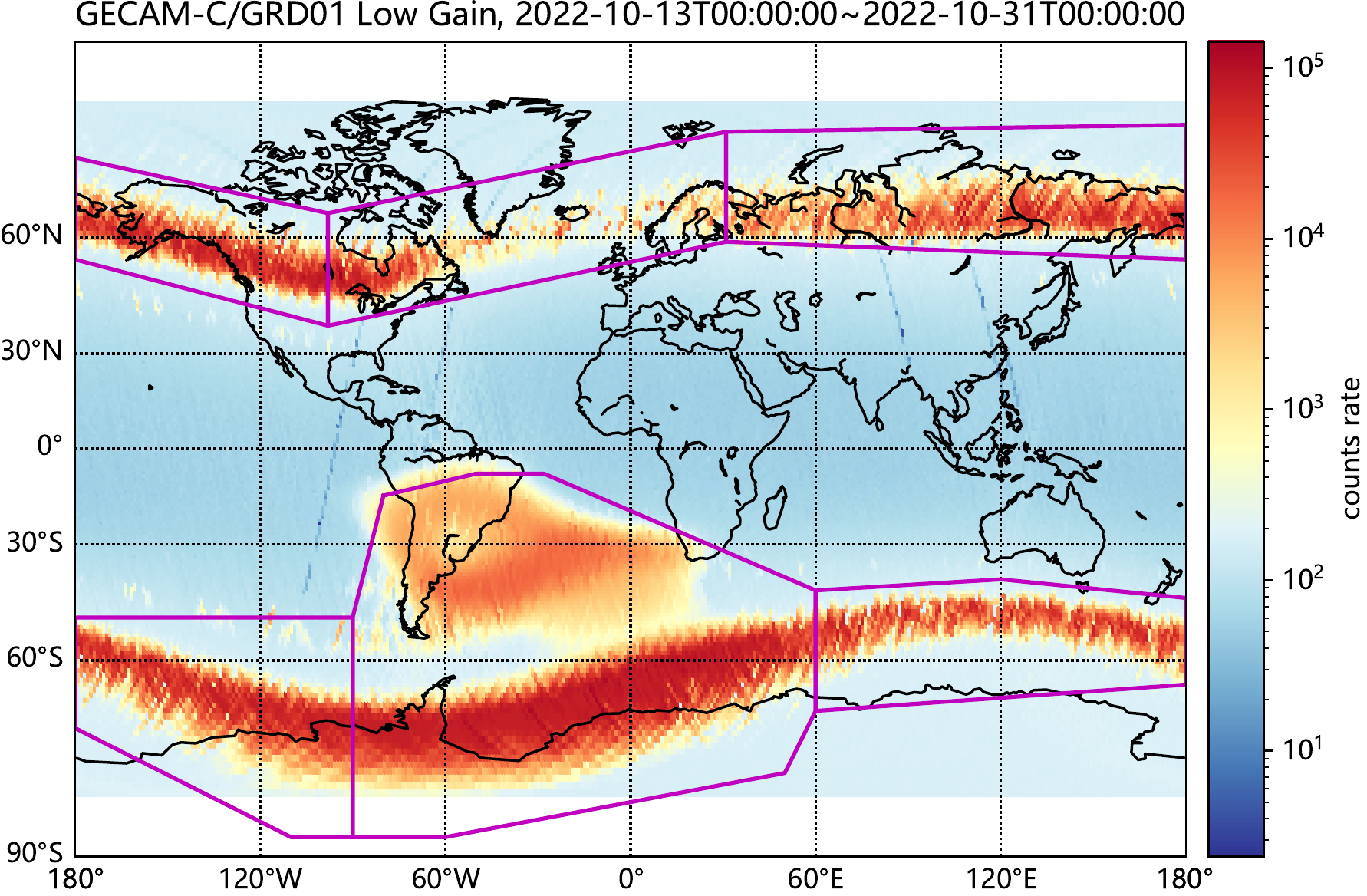}
  \hspace{0.1cm}
  \includegraphics[width=6.4 cm]{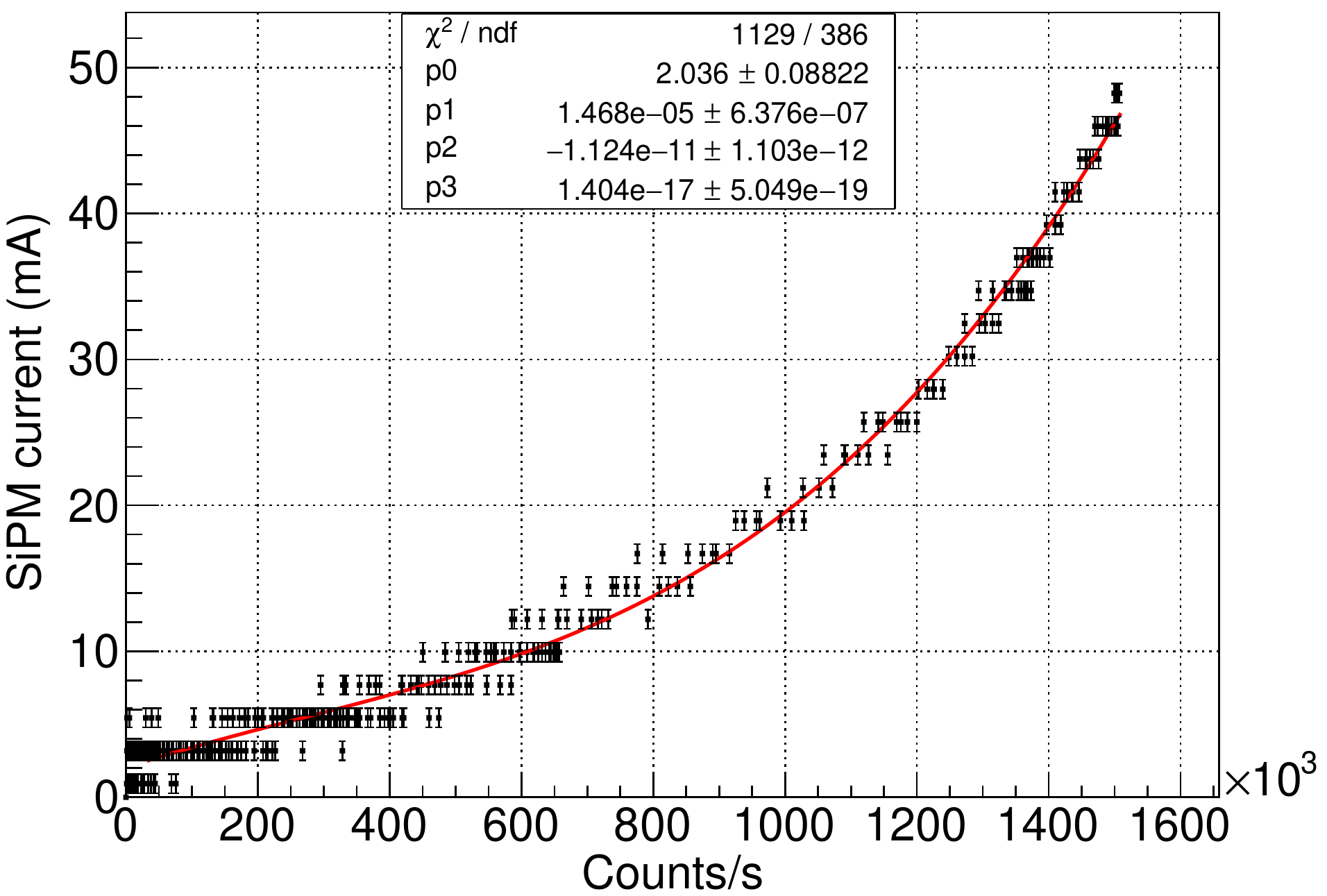}
  \caption{Left panel: Counts rate distribution with satellite position. Right panel: Total SiPM current Vs counts rate.}\label{Fig13}
  \vspace{0cm}
\end{figure}
\par Fig. \ref{Fig13} left panel shows the counts rate distribution with satellite position which is measured by GRD01. Purple polygons are the so-called high latitude area we defined for GECAM-C, where the background rate is much higher than other part of orbit. Fig. \ref{Fig13} right panel shows that the high counts rate of proton and electron will cause a significant SiPM current increase which is unfavorable for SiPM operation. Thus we designed a dedicated operation strategy for GECAM-C which is to turn GRD01 and CPD01 on all the time to monitor the in-flight background while other detectors are turned off in high counts rate area (i.e. purple polygons).
\par
\begin{figure}[htbp]
  \centering
  \includegraphics[width=6.5 cm]{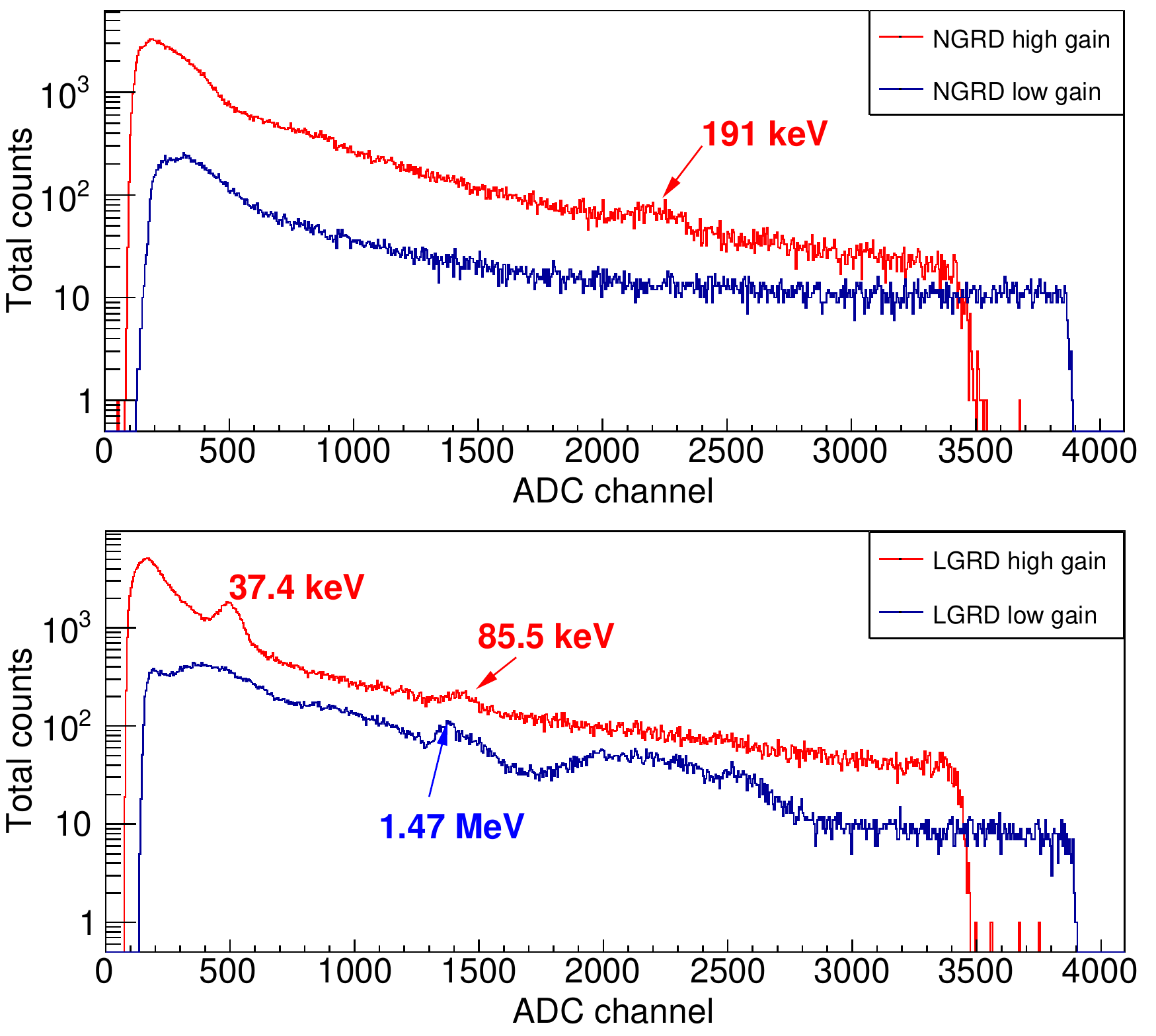}
  \hspace{0.1cm}
  \includegraphics[width=7 cm]{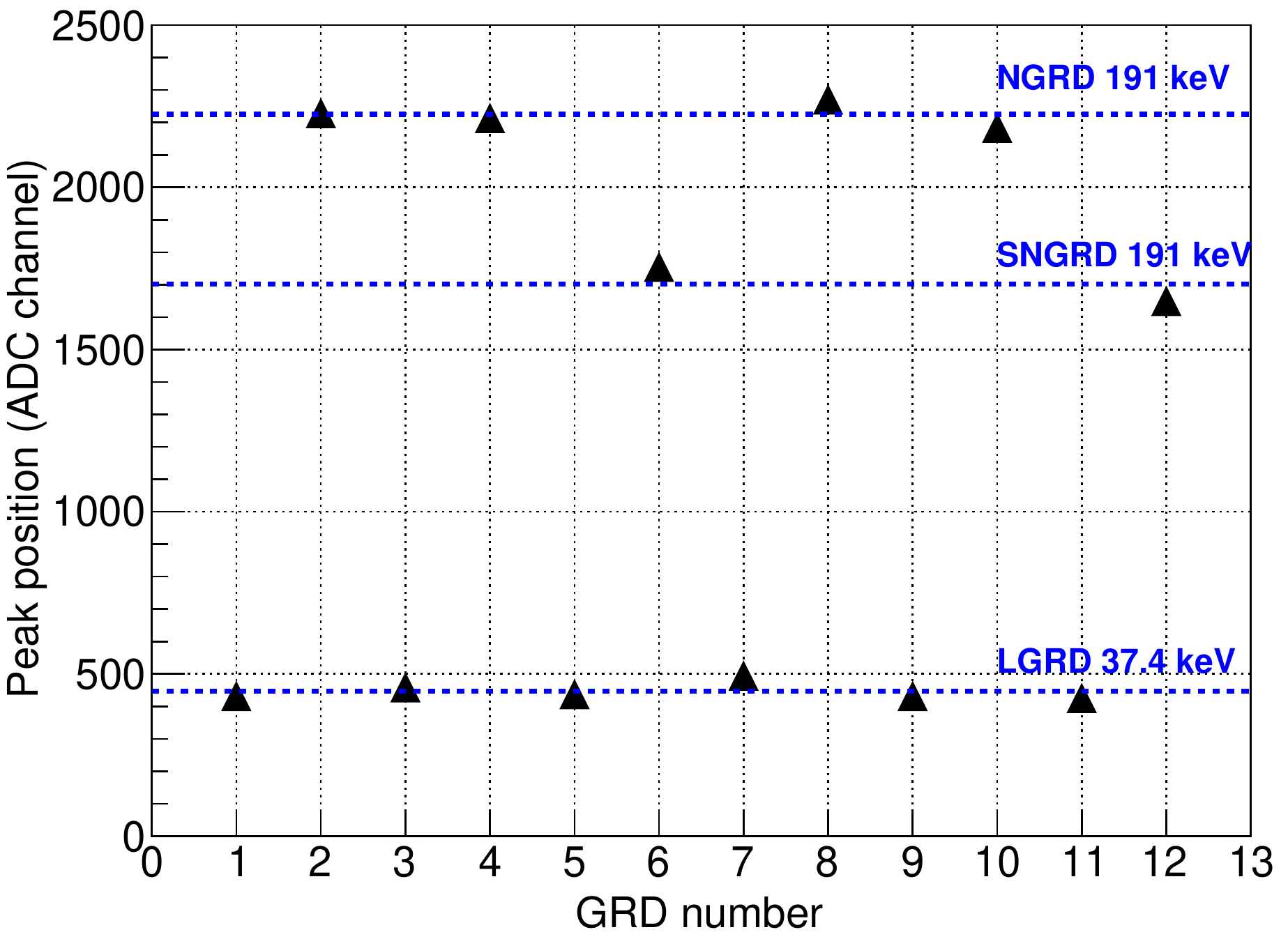}
  \caption{Left panel: Measured in-flight background spectrum of GRD. Right panel: Alignment of energy gain of GRDs with the peak position of in-flight energy lines.}\label{Fig14}
  \vspace{0cm}
\end{figure}

 Fig. \ref{Fig14} left panel shows the measured in-flight background in low counts rate area. NGRDs and LGRDs have different intrinsic activities and activated energy lines. For NGRD, NaI does not have intrinsic activity energy line and the activated 191 keV energy line \cite{NaI_BK} can be observed in high gain channel. For LGRD, the 37.4 keV energy line in high gain channel and 1470 keV energy line in low gain channel are the intrinsic activities of LaBr$_{3}$ and 85.5 keV energy line in high gain channel is the activated energy line. Fig. \ref{Fig14} right panel shows the alignment of line peak positions (i.e. the energy gain) of all GRDs. The gain of LGRDs and NGRDs were adjusted according to the peak positions of 37.4 keV and 191 keV energy lines, respectively. With these in-flight characteristic lines and the detailed E-C relation already measured on ground, we can derive the in-flight E-C relation for GRDs.

\subsection{In-flight performance of SiPM}
\par Although SiPM-based GRD have many advantages compared to PMT-based detectors, there still remain questions on whether SiPM-based GRD can meet the performance requirements during the full period of operation considering the radiation damage issues of SiPM. There has been an increasing interest in assessing the SiPM performance evolution under irradiation, especially in the in-orbit data \cite{SiPMRadiationDamage} \cite{GRIDRadiationDamage}. The in-orbit irradiation will cause SiPM dark current growth and the low energy detection capability deterioration of GRD. Based on the precious SiPM operation data of GECAM-B, we made several important improvements for GECAM-C which are listed in Table \ref{InFlightEvolution}.
\par As shown in Fig. \ref{Fig16}, the SiPM current growth of GECAM-B is much more obvious than GECAM-C. The quick SiPM growth of GECAM-B is caused by the harsh irradiation environment, higher operating temperature and relatively high SiPM bias voltage compared to GECAM-C. The simulated irradiation dose of GECAM-B and GECAM-C is 724.8 and 95.8 rad per year, respectively. Because they have different orbit, the satellite has been working in a higher temperature than GECAM-C. The unexpected SiPM current growth of GECAM-B forced us to adjust the SiPM bias voltage several times but it still works in a higher voltage than GECAM-C, due to the limited adjustment range of the bias voltage. The current growth rate of GECAM-B is 0.429 $\mu$A/day under an average SiPM bias voltage of 28.16 V. The optimized SiPM bias voltage makes the GECAM-C have a much lower SiPM current growth rate of 0.046 $\mu$A/day under an average SiPM bias voltage of 26 V.

\begin{figure}[htbp]
  \centering
  \setlength{\belowcaptionskip}{0.1cm}
  \includegraphics[width=12 cm]{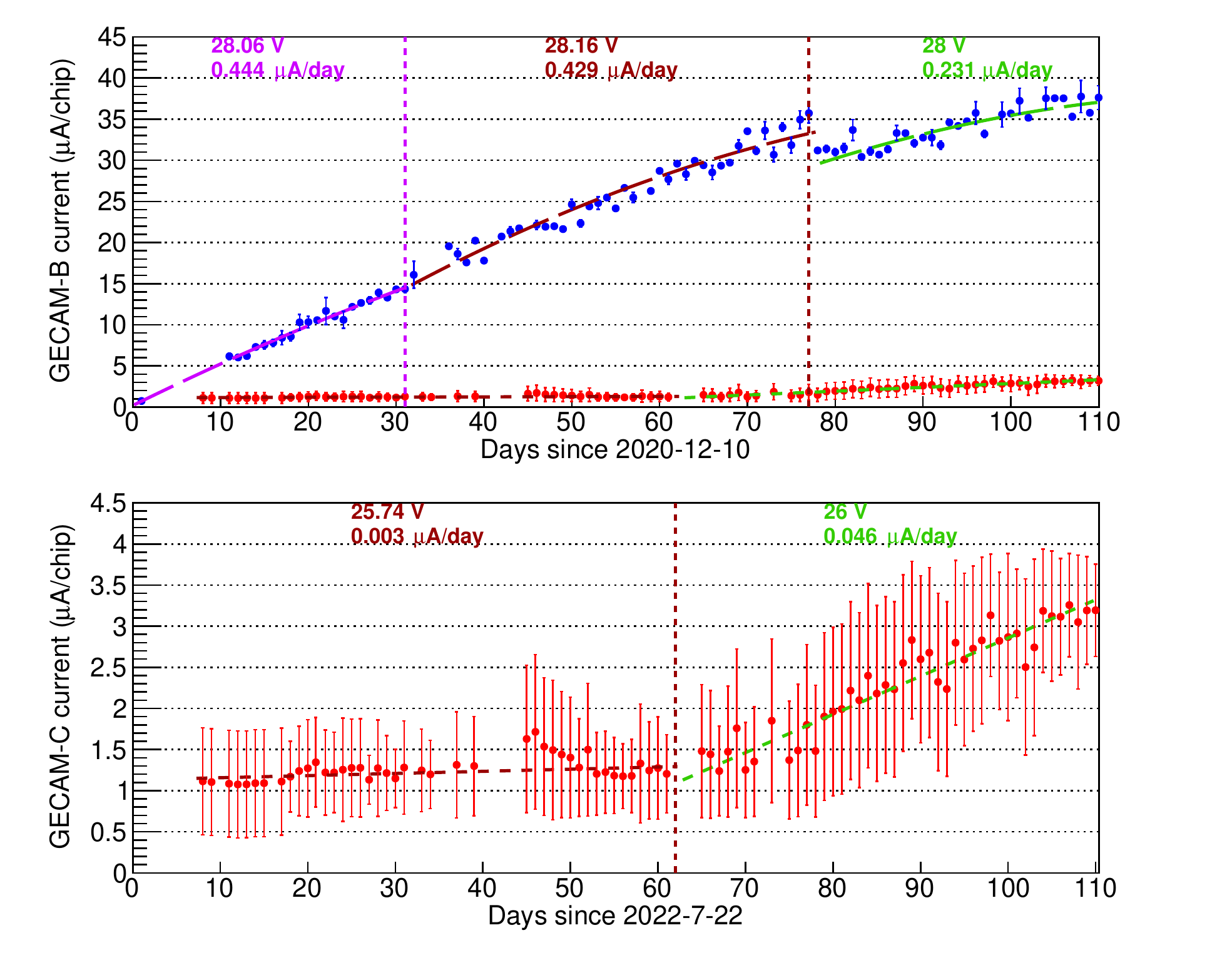}
  \caption{Top panel: SiPM current evolution of GECAM-B (blue markers). The average voltage at -25 $^{\circ}$C in the SiPM bias voltage look-up table and current growth rate are labeled in the plot. GECAM-C data (red markers) are also plot for comparison. Bottom panel: SiPM current evolution of GECAM-C. The corresponding SiPM voltage and current growth rate are also labeled in figure.}\label{Fig15}
\end{figure}
\par Low current growth rate ensures a steady SiPM performance and long operating lifetime. Fig. \ref{Fig16} shows the comparison of SiPM noise evolution between GECAM-B and GECAM-C. For GECAM-B, the SiPM noise is within 4 keV just after launch and then rapidly increased in a growth rate of 0.041 keV/day in the first month and 0.1 keV/day in day 31 to 78. The noise growth comes from the temperature increase of the satellite platform and high SiPM bias voltage. When the operating temperature and SiPM bias voltage decreased, the SiPM noise growth rate became stable.
\begin{figure}[htbp]
  \centering
  \setlength{\belowcaptionskip}{0.1cm}
  \includegraphics[width=12 cm]{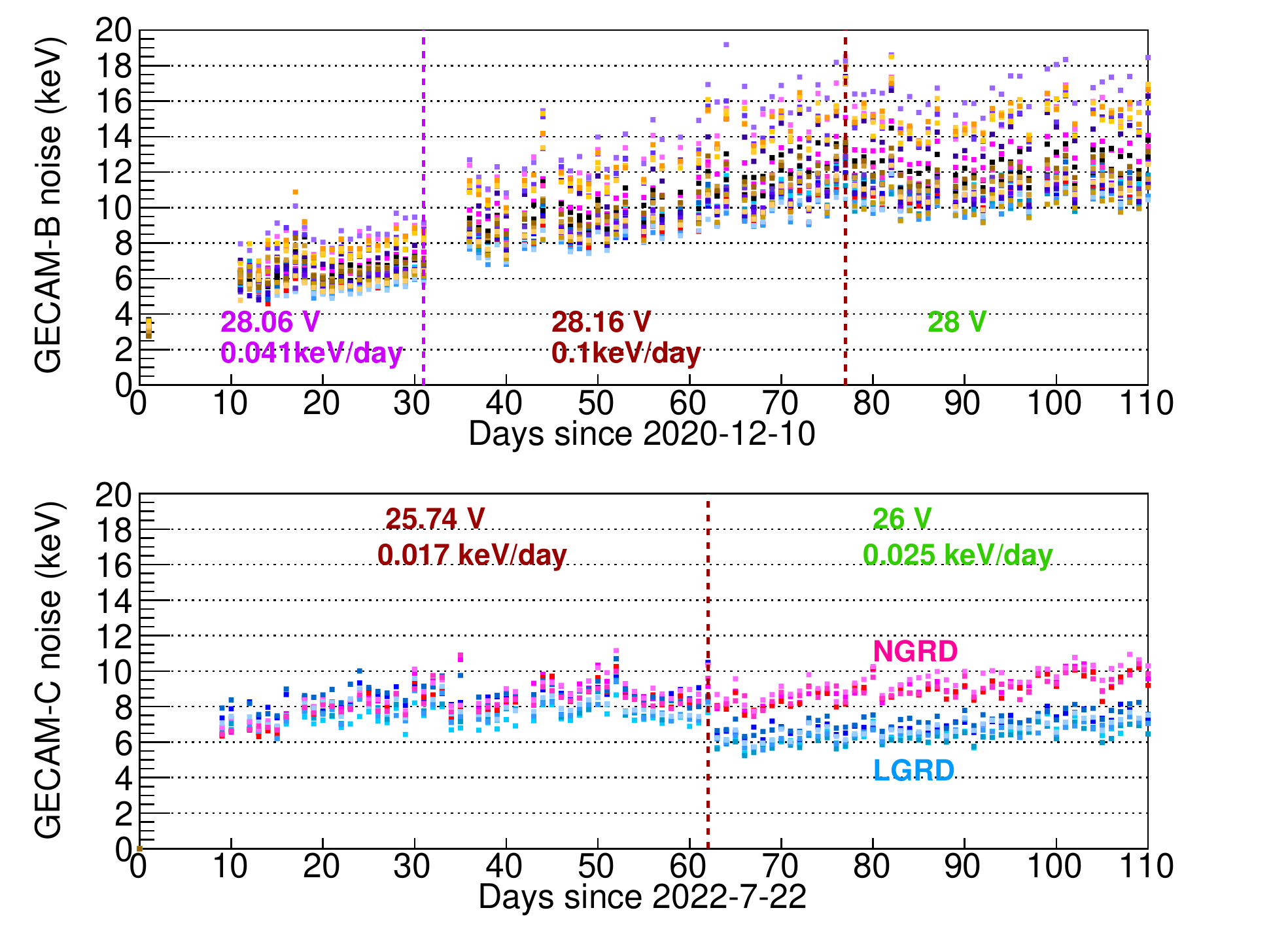}
  \caption{Comparison of SiPM noise evolution. The SiPM noise of all GRDs are labeled with different colors. For GECAM-C, the date of NGRDs and LGRDs are labeled with reddish and bluish colors, respectively.}\label{Fig16}
\end{figure}
\par For GECAM-C, the SiPM bias look-up table was updated on day 63 to achieve E-C relation consistency among GRD detectors. The average SiPM bias voltage of NGRD is higher than LGRD which results in a little bit higher noise growth rate. The SiPM noise growth rate of GECAM-C is slow (0.025 keV/day ) which is benficial for a long operating lifetime. These results show that lower orbit height (less irradiation dose), lower SiPM bias voltage and lower operation temperature could effectively slow the performance deterioration of SiPM. More data will be accumulated to study the SiPM irradiation damage model based on the data of GECAM instruments. These results will provide important reference for incoming transient monitor projects such as POLAR2 \cite{POLAR2} which also utilizes the SiPM.
\begin{table}[H]
\centering
\caption{Specification of GRD in-flight performance. The current and noise growth rate is measured under
-20$\backsim$-15$^{\circ}$C for GECAM-B and -24$\backsim$-21 $^{\circ}$C for GECAM-C.
}
\begin{tabular}{|c|c|c|c|}
\hline
Parameters	                    &GECAM-B	                   &GECAM-C                      \\ \hline
Track height and inclination	&595.5 km, 29$^{\circ}$	       &500 km, 97.4$^{\circ}$        \\ \hline
Operating temperature           &-25 $\backsim$ -5 $^{\circ}$C &-25 $\backsim$ -15 $^{\circ}$C \\ \hline
SiPM bias voltage	            &27.5$\backsim$29.5 V          &24.5$\backsim$26.5 V         \\ \hline
Simulated irradiation dose	    &724.8 rad/year	               &95.8 rad/year                \\ \hline
Current growth rate per chip  &0.429 $\mu$A@28.16 V/day	   &0.046 $\mu$A@26 V/day       \\ \hline
GRD noise growth rate	        &0.1 keV@28.16 V/day	       &0.025 keV@26 V/day          \\ \hline
\end{tabular}
\label{InFlightEvolution}
\end{table}
\section{Conclusions and future work}
\par The SiPM-based gamma-ray detector of GECAM-C inherited the detector technology from GECAM-A and GECAM-B and several modifications have been made according to the scientific and engineering requirements. In the on-ground tests, dedicated calibration tests were performed to study the GRD energy response by using the X-ray beam facility. The GRD temperature dependence was tested and proper temperature compensation approach were determined. After all GRDs were assembled on the GECAM-C satellite, the final ground calibration tests were done in China’s Jiuquan Satellite Launch Center. After GECAM-C was successfully launched, the in-flight background analysis of GRD indicates that in-flight energy lines can be used for calibration. Then the in-flight energy gain calibration and detector response matrix can be derived. During the early days of the commissioning phase, the performance of GECAM-C has been intensively optimized, which pave the way for GECAM-C to accurately observe the exceptional bright GRB 221009A.
\par The long-term evolution of GECAM-C GRD performance will be closely monitored and studied, especially the degration of SiPM caused by irradiation damage on-orbit. Comparison study between these GECAM instruments (GECAM-A, GECAM-B and GECAM-C) will provide unique information for SiPM-based detection technology.

\section*{Acknowledgements}
\par \emph{We would like to express our appreciation to the staff of the Key Laboratory of Particle Astrophysics, Center for Space Science and Applied Research, Shandong Institute of aerospace electronic technology and National Institute of Metrology who offer great help in the development and tests of GECAM-C. This work is supported by the National Key R\&D Program of China (2021YFA0718500). This research is supported by the National Science Foundation for Young Scientists of China, Grant No.12203053 , Strategic Priority Research Program of Chinese Academy of Sciences, Grant No.XDA 15360102 and National Natural Science Foundation of China (12173038, 12273042).}

\bibliography{mybibfile}

\end{document}